\documentclass[twocolumn,amsmath,amssymb,prd,showpacs]{revtex4}
\usepackage{psfig}

\def\al{\alpha}
\def\be{\beta}
\def\ga{\gamma}
\def\de{\delta}
\def\ep{\epsilon}

\def\et{\eta}
\def\th{\theta}

\def\la{\lambda}

\def\rh{\rho}

\def\si{\sigma}

\def\ta{\tau}

\def\ph{\phi}
\def\vp{\varphi}
\def\ch{\chi}

\def\om{\omega}
\def\Ga{\Gamma}
\def\De{\Delta}
\def\Th{\Theta}
\def\La{\Lambda}

\def\Ph{\Phi}

\def\Om{\Omega}
\def\mn{{\mu\nu}}

\def\cl{{\cal L}}

\def\fr#1#2{{{#1} \over {#2}}}
\def\frac#1#2{\textstyle{{{#1} \over {#2}}}}

\def\prt{\partial}
\def\vev#1{\langle {#1}\rangle}

\def\half{{\textstyle{1\over 2}}}
\def\lsim{\mathrel{\rlap{\lower4pt\hbox{\hskip1pt$\sim$}}
    \raise1pt\hbox{$<$}}}
\def\gsim{\mathrel{\rlap{\lower4pt\hbox{\hskip1pt$\sim$}}
    \raise1pt\hbox{$>$}}}
\def\Re{\hbox{Re}\,}
\def\Im{\hbox{Im}\,}

\def\etal {{\it et al.}}
\newcommand{\beq}{\begin{equation}}
\newcommand{\eeq}{\end{equation}}
\newcommand{\bea}{\begin{eqnarray}}
\newcommand{\eea}{\end{eqnarray}}
\newcommand{\bse}{\begin{subequations}}
\newcommand{\ese}{\end{subequations}}
\newcommand{\rf}[1]{(\ref{#1})}

\def\to{\rightarrow}

\def\mix{\leftrightarrow}

\def\nub{\bar\nu}
\def\vp{\vec p}
\def\cmat{{\cal C}}
\def\cH{{\cal H}}
\def\heff{h_{\rm eff}}
\def\Ueff{U_{\rm eff}}
\def\tu{\widetilde U}
\def\bu{\overline U}
\def\ml{m_l}
\def\mt{\widetilde m^2}
\def\gt{\tilde g} 
\def\Ht{\widetilde H} 
\def\AA{{A'}}
\def\BB{{B'}}

\def\aa{{a'}}
\def\bb{{b'}}

\def\aaa{{\hat a'}}
\def\bbb{{\hat b'}}
\def\ring#1{{\mathaccent'27 #1}}
\def\cri{\ring{c}}
\def\ari{\ring{a}}
\def\mri{\ring{m}}
\def\mem{\mri}
\def\aem{\ari}
\def\amt{\ari'}

\def\cmt{\cri}
\def\Gc{\check{g}}
\def\a3em{\check{a}}
\def\cee{\cri}

\begin{document}
\title{Lorentz and CPT violation in neutrinos}
\author{V.\ Alan Kosteleck\'y and Matthew Mewes}
\affiliation{Physics Department, Indiana University, 
         Bloomington, IN 47405, U.S.A.}
\date{IUHET 459, August 2003} 

\begin{abstract}
A general formalism is presented for violations 
of Lorentz and CPT symmetry in the neutrino sector.
The effective hamiltonian for neutrino propagation 
in the presence of Lorentz and CPT violation is derived,
and its properties are studied.
Possible definitive signals in existing and future 
neutrino-oscillation experiments are discussed.
Among the predictions are direction-dependent effects,
including neutrino-antineutrino mixing,
sidereal and annual variations, 
and compass asymmetries.
Other consequences of Lorentz and CPT violation
involve unconventional energy dependences
in oscillation lengths and mixing angles.
A variety of simple models both with and without neutrino masses 
are developed to illustrate key physical effects.
The attainable sensitivities 
to coefficients for Lorentz violation in the Standard-Model Extension
are estimated for various types of experiments.
Many experiments have potential sensitivity
to Planck-suppressed effects,
comparable to the best tests in other sectors.
The lack of existing experimental constraints,
the wide range of available coefficient space,
and the variety of novel effects
imply that some or perhaps even all of the existing data 
on neutrino oscillations might be due to Lorentz and CPT violation.

\end{abstract}


\maketitle


\section{Introduction}

The minimal Standard Model (SM) of particle physics
offers a successful description of most processes in Nature
but leaves unresolved several experimental and theoretical issues.
On the experimental front,
observations of neutrino oscillations
have accumulated convincing evidence
that the description of physical properties of neutrinos
requires modification of the neutrino sector 
in the minimal SM.
Most experimental results to date can be described theoretically
by adding neutrino masses to the minimal SM,
but a complete understanding of the existing data
awaits further experimentation.
On the theoretical front,
the SM is expected to be the low-energy limit
of a more fundamental theory
that unifies quantum physics and gravity at the Planck scale,
$m_P \simeq 10^{19}$ GeV.
Direct measurements at this energy scale are impractical,
but suppressed low-energy signatures
from the anticipated new physics
might be detectable in sensitive existing experiments.

In this work,
we address both these topics by studying effects on the neutrino sector
of relativity violations, 
a promising class of Planck-scale signals.
These violations might arise through the breaking of Lorentz symmetry
and perhaps also the breaking of CPT symmetry
\cite{cpt01}.
Since the SM is known to provide a successful description
of most physics at low energies compared to the Planck scale,
any such signals must appear at low energies 
in the form of an effective quantum field theory
containing the SM.
The general effective quantum field theory constructed from the SM
and allowing arbitrary coordinate-independent Lorentz violation
is called the Standard-Model Extension (SME)
\cite{ck}.
It provides a link to the Planck scale 
through operators of nonrenormalizable dimension
\cite{kpo,kle}.
Since CPT violation implies Lorentz violation
\cite{owg},
this theory also allows for general CPT breaking.
The SME therefore provides a realistic theoretical basis
for studies of Lorentz violation,
with or without CPT breaking.

The lagrangian of the SME consists of the usual SM lagrangian
supplemented by all possible terms that can be constructed
with SM fields and that introduce violations of Lorentz symmetry.
The additional terms
have the form of Lorentz-violating operators
coupled to coefficients with Lorentz indices,
and they could arise in a variety of ways.
One generic and elegant mechanism is spontaneous Lorentz violation,
proposed first in string theory and field theories with gravity
\cite{ks}
and then generalized to include CPT violation
\cite{kp}.
Another popular framework for Lorentz violation is
noncommutative field theory,
in which realistic models form a subset of the SME
involving operators of nonrenormalizable dimension
\cite{ncqed}.
Other proposed sources of Lorentz and CPT violation include
various non-string approaches to quantum gravity
\cite{qg},
random dynamics
\cite{fn}
and multiverses
\cite{bj}.
Planck-scale sensitivity
to the coefficients for Lorentz violation in the SME
has been achieved in various experiments,
including ones with
mesons \cite{hadronexpt,kpo,hadronth},
baryons \cite{ccexpt,spaceexpt,cane},
electrons \cite{eexpt,eexpt2},
photons \cite{photonexpt,photonth,cavexpt,km},
and muons \cite{muons}.
However,
no experiments to date have measured 
neutrino-sector coefficients for Lorentz violation.

Here,
we explore neutrino behavior 
in the presence of Lorentz and CPT violation 
using the SME framework.
The original proposal for Lorentz and CPT violation in neutrinos
\cite{ck}
has since been followed by several theoretical investigations
within the context of the SME
\cite{fc1,fc2,fc3,fc4,fc5,nu},
most of which have chosen to restrict attention 
to a small number of coefficients.
A comprehensive theoretical study of Lorentz and CPT violation
in neutrinos has been lacking.
The present work partially fills this gap by applying 
the ideas of the SME to a general neutrino sector
with all possible couplings of left- and right-handed neutrinos
and with sterile neutrinos.
We concentrate mostly on Lorentz-violating operators 
of renormalizable dimension,
which dominate the low-energy physics in typical theories,
but some generic consequences of Lorentz-violating operators 
of nonrenormalizable dimension are also considered
\cite{kpo,kle,bef}. 
The effective hamiltonian describing free neutrino propagation 
is obtained,
and its implications are studied.
The formalism presented in this work
thereby provides a general theoretical basis
for future studies of Lorentz and CPT violation in neutrinos.
We also illustrate various key physical ideas 
of Lorentz and CPT violation through simple models, 
and we discuss experimental signals.
Our primary focus here is on oscillation data
\cite{pdg},
but the formalism is applicable 
also to other types of experiments including 
direct mass searches
\cite{katrin},
neutrinoless double-beta decay
\cite{bbdecay},
and supernova neutrinos
\cite{sn1987a}.

Several features of Lorentz and CPT violation that we uncover 
are common to other sectors of the SME,
including unconventional energy dependence
and dependence on the direction of propagation.
We also find that Lorentz-violating neutrino-antineutrino mixing
with lepton-number violation 
naturally arises from Majorana-like couplings.
These features lead to several unique signals 
for Lorentz and CPT violation.
For example,
the direction dependence potentially generates 
sidereal variations in terrestrial experiments as the Earth rotates, 
annual variations in solar-neutrino properties,
and intrinsic differences in neutrino flux 
from different points on the compass or different angular heights
at the location of the detector.
The unconventional energy dependence produces a variety
of interesting potential signals, 
including resonances in the vacuum
\cite{fc2,nu} 
as well as the usual MSW resonances in matter
\cite{msw}.

Experiments producing evidence for neutrino oscillations 
to date include 
atmospheric-neutrino experiments
\cite{sk},
solar-neutrino experiments
\cite{homestake,gallex,gno,sage,sksol,sno},
reactor experiments
\cite{kamland},
and accelerator-based experiments
\cite{lsnd,k2k}.
Most current data are consistent 
with the introduction of three massive-neutrino states,
usually attributed to GUT-scale physics.
However, 
as we demonstrate in this work,
the possibility remains that the observed neutrino oscillations 
may be due at least in part and conceivably even entirely
to Lorentz and CPT violation from the Planck scale.
In any event,
experiments designed to test neutrino mass 
are also well suited for tests of Lorentz and CPT invariance,
and they have the potential to produce the first measurements 
of violations of these fundamental symmetries,
signaling possible Planck-scale physics.

The organization of this paper is as follows.
Section \ref{theory} presents the basic theory and definitions,
obtaining the effective hamiltonian for neutrino propagation 
and discussing its properties.
Issues of experimental sensitivities 
and possible constraints from experiments in other sectors 
are considered in Section \ref{sens}.
Certain key features of neutrino behavior in the presence of 
Lorentz and CPT violation are illustrated 
in the sample models of Section \ref{models}.
Some remarks about both generic and experiment-specific predictions
are provided in Section \ref{disc}.
Throughout,
we follow the notation and conventions 
of Refs.\ \cite{ck,kle}.


\section{Theory}\label{theory}

\subsection{Basics}\label{basics}

Our starting point is a general theory describing $N$ neutrino species.
The theory is assumed to include all possible 
Majorana- and Dirac-type couplings of left- and right-handed neutrinos,
including Lorentz- and CPT-violating ones. 
The neutrino sector of the minimal SME is therefore included,
along with other terms such as those involving right-handed neutrinos.

We denote the neutrino fields by the set of Dirac spinors
\{$\nu_e$, $\nu_\mu$, $\nu_\ta$,\ldots \}
and their charge conjugates by
\{$\nu_{e^C}\equiv \nu^C_e$,
$\nu_{\mu^C}\equiv \nu^C_\mu$,
$\nu_{\ta^C}\equiv \nu^C_\ta$,\ldots \},
where charge conjugation of a Dirac spinor
is defined as usual:
$\nu^C_a \equiv C\nub^T_a$.
By definition,
active neutrinos are detected via
weak interactions with left-handed
components of \{$\nu_e$, $\nu_\mu$, $\nu_\ta$\}.
Complications may arise in the full SME,
where Lorentz-violating terms alter these interactions
and can modify the detection process.
However, 
such modifications are expected to be tiny and
well beyond the sensitivity of current experiments.
In contrast, 
propagation effects can become appreciable for large baselines.
We therefore focus in this work
on solutions to the Lorentz-violating equations of motion
that describe free propagation of the $N$ neutrino species.

It is convenient to place all the fields and their conjugates
into a single object $\nu_A$,
where the index $A$ ranges over the $2N$ possibilities
$\{e,\mu,\ta,\ldots, e^C,\mu^C,\ta^C,\ldots \}$.
This setup allows us to write the equations of motion
in a form analogous to the Lorentz-violating QED extension 
\cite{ck,kle},
and it can readily accommodate Dirac, Majorana, 
or more general types of neutrinos.
Our explicit analysis in this section is performed 
under the assumption that Lorentz-violating operators 
of renormalizable dimension dominate the low-energy physics.
Then,
the general equations of motion for free propagation
can be written as a first-order differential operator acting on
the object $\nu_A$:
\beq
(i\Ga^\nu_{AB}\prt_\nu-M_{AB})\nu_B=0 .
\label{de}
\eeq
Here,
each constant quantity $\Ga^\mu_{AB}$, $M_{AB}$
is also a $4 \times 4$ matrix in spinor space.
Note that the usual equations of motion 
for Dirac and Majorana neutrinos 
are special cases of this equation.

The matrices $\Ga^\mu_{AB}$ and $M_{AB}$ can be decomposed
using the basis of $\ga$ matrices.
We define 
\bea
\Ga^\nu_{AB} &\equiv&
\ga^\nu \de_{AB}
+ c^\mn_{AB}\ga_\mu
+ d^\mn_{AB}\ga_5\ga_\mu
\nonumber \\
&& + e^\nu_{AB}
+ if^\nu_{AB}\ga_5
+ \half g^{\la\mn}_{AB}\si_{\la\mu} ,
\nonumber\\
M_{AB} &\equiv&
m_{AB}
+im_{5AB}\ga_5
\nonumber \\
&& + a^\mu_{AB}\ga_\mu
+ b^\mu_{AB}\ga_5\ga_\mu
+ \half H^\mn_{AB}\si_\mn .
\label{GaM}\eea
In these equations,
the masses $m$ and $m_5$ are Lorentz and CPT conserving.
The coefficients $c$, $d$, $H$
are CPT conserving but Lorentz violating,
while $a$, $b$, $e$, $f$, $g$
are both CPT and Lorentz violating.
Requiring hermiticity of the theory imposes the conditions 
$\Ga^\nu_{AB} = \ga^0(\Ga^\nu_{BA})^\dag\ga^0$
and $M_{AB} = \ga^0(M_{BA})^\dag\ga^0$,
which implies all coefficients are hermitian in generation space.

The above construction carries some redundancies 
that stem from the interdependence of $\nu$ and $\nu^C$.
This implies certain symmetries for $\Ga^\nu$ and $M$.
Note first that charge conjugation can be written
as a linear transformation on $\nu_A$:
$\nu^C_A = \cmat_{AB} \nu_B$,
where $\cal C$ is the symmetric matrix
with nonzero elements
$\cmat_{e e^C}
=\cmat_{\mu \mu^C}
=\cmat_{\ta \ta^C}=\cdots=1$.
Then,
in terms of $\cal C$ and the spinor matrix $C$,
the interdependence of $\nu$ and $\nu^C$ implies the relations
\bea
\Ga^\nu_{AB}&=&
-\cmat_{AC}\cmat_{BD}C(\Ga^\nu_{DC})^TC^{-1} ,
\nonumber\\
M_{AB}&=&\cmat_{AC}\cmat_{BD}C(M_{DC})^TC^{-1} ,
\label{GaMC}\eea
where the transpose $T$ acts in spinor space.
Suppressing generation indices,
this translates to
\beq
\begin{array}{rclrcl}
c^\mn&=&\cmat(c^\mn)^T\cmat , &
m&=&\cmat (m)^T \cmat , \\
d^\mn&=&-\cmat (d^\mn)^T \cmat , &
m_{5}&=& \cmat (m_5)^T   \cmat , \\
e^\nu&=&-\cmat (e^\nu)^T \cmat , &
a^\nu&=&-\cmat (a^\nu)^T \cmat , \\
f^\nu&=&-\cmat (f^\nu)^T \cmat  , &
b^\nu&=& \cmat (b^\nu)^T \cmat , \\
g^{\la\mn}&=&\cmat (g^{\la\mn})^T \cmat , &
H^\mn&=&-\cmat (H^\mn)^T \cmat ,
\end{array}
\label{C}
\eeq
where now the transpose $T$ acts in generation space.
Note that the overall sign in the above equations are chosen
to match their derivation within the conventional
lagrangian formalism involving anticommuting fermion fields.

Equation \rf{de} provides a basis
for a general Lorentz- and CPT-violating
relativistic quantum mechanics of freely propagating neutrinos.
However,
the unconventional time-derivative term complicates
the construction of the corresponding hamiltonian.
This difficulty also arises in the minimal QED extension,
but it may be overcome 
\cite{kle}
if there exists 
a nonsingular matrix $A$ satisfying the relationship
$A^\dag\ga^0\Ga^0A=1$.
The field redefinition
$\nu_A = A_{AB} \ch_B$
then allows the equations of motion \rf{de} 
to be written as
$(i\de_{AB}\prt_0-\cH_{AB}) \ch_B = 0$,
where the hamiltonian is given by
$\cH=-A^\dag\ga^0(i\Ga^j\prt_j-M)A$.

Denoting $\de\Ga^\nu$ and $\de M$
as the Lorentz-violating portions
of $\Ga^\nu$ and $M$,
and under the reasonable assumption that $|\de\Ga^0| < 1$,
a satisfactory field redefinition is
given by the power series
$A=(1+\ga^0\de\Ga^0)^{-1/2}=1-\half\ga^0\de\Ga^0+\cdots$.
Separating the hamiltonian $\cH$ into a
Lorentz-conserving part $\cH_0$
and a Lorentz-violating part $\de \cH$,
which we assume is small relative to $\cH_0$,
we can use the above expression for $A$ 
to obtain an expansion of $\de \cH$
in terms of $\cH_0$ and coefficients for Lorentz violation.
Explicitly,
at leading order in coefficients for Lorentz violation,
we obtain 
\beq
\de\cH = -\half(\ga^0\de\Ga^0\cH_0+\cH_0\ga^0\de\Ga^0)
-\ga^0(i\de\Ga^j\prt_j-\de M) . 
\label{dH}
\eeq
This expression is therefore the basis for
a general study of leading-order Lorentz and CPT violation 
in the neutrino sector. 

At this stage,
prior to beginning our study of Eq.\ \rf{dH}, 
it is useful to review the properties
of the Lorentz-conserving hamiltonian
\cite{nuphysics,nuphysics2}
\beq
\cH_0=-\ga^0(i\ga^j\prt_j-M_0).
\eeq
The Lorentz-conserving dynamics
is completely determined by the mass matrix $M_0$, 
which in its general form can be written
\beq
M_0=m+im_5\ga_5 =m_L P_L+m_R P_R\ ,
\eeq
with $m_R=(m_L)^\dag=m+im_5$ and
$P_L=\half(1-\ga_5), P_R=\half(1+\ga_5)$.
The components of the matrix $m_R=m^\dag_L$
can be identified with Dirac- or
Majorana-type masses by separating $m_R$ 
into four $N \times N$ submatrices.
It is often encountered in
the form of the symmetric matrix
\beq
m_R \cmat=
\left(\begin{array}{cc}
L & D \\
D^T & R
\end{array}\right) .
\label{usumass}
\eeq
The matrices $R$ and $L$ are the
right- and left-handed Majorana-mass matrices,
while $D$ is the Dirac-mass matrix.
In general, $R$, $L$ and $D$
are complex matrices
restricted only by the requirement that
$R$ and $L$ are symmetric.
Note that a left-handed Majorana coupling
is incompatible with electroweak-gauge invariance.
In contrast, Dirac and right-handed Majorana
couplings can preserve the usual gauge invariance.

It is always possible to find a basis
in which the mass matrix $M_0$ is diagonal.
Labeling the fields in this basis by $\ch_\AA$,
where $\AA  = 1,\ldots,2N$,
then the unitary transformation
relating the two bases can be written as 
\beq
U_{\AA A}=V_{\AA A} P_L
+(V\cmat)^*_{\AA A} P_R,
\label{U}
\eeq
where $V$ is a $2N \times 2N$
unitary matrix.
Here, 
it is understood that
$U_{\AA A}$ carries spinor indices that
have been suppressed.
In the new basis, the mass matrix
$m_{L\AA \BB}
=m_{R\AA \BB}
=m_{(\AA )}\de_{\AA \BB}$
is diagonal with real nonnegative entries.
The neutrinos
$\ch_\AA=\ch^C_\AA
=V_{\AA A} P_L \ch_A
+V^*_{\AA A} P_R \ch^C_A$
are Majorana particles,
regardless of the form of $M_0$.

\subsection{Effective hamiltonian}\label{phenom}

The discussion above applies to an arbitrary number 
of neutrino species and an arbitrary mass spectrum.
Since a general treatment is rather cumbersome,
we restrict attention in what follows 
to the minimal physically reasonable extension with $N=3$.
For definiteness,
we also assume a standard seesaw mechanism
\cite{seesaw}
with the components of $R$ much larger than those of $D$ or $L$.
This mechanism suppresses the propagation of
right-handed neutrinos,
so the analysis below also contains 
other Lorentz- and CPT-violating scenarios dominated 
by light or massless left-handed neutrinos,
including the minimal SME.

Ordering the masses $m_{(\AA )}$ from smallest to largest,
we assume that $m_{(1)}$, $m_{(2)}$, $m_{(3)}$
are small compared to the neutrino energies and possibly zero,
and that the remaining masses $m_{(4)}$, $m_{(5)}$, $m_{(6)}$
are large with the corresponding energy eigenstates 
kinematically forbidden.
In this situation the submatrix $V_{\aa a}$, 
where $a=e,\mu,\ta$ and $\aa =1,2,3$, 
is approximately unitary.

To aid in solving the equations of motion, 
we define
\bea
\ch_A(t;\vec x)
&=&\int \fr{d^3p}{(2\pi)^3}\ch_A(t;\vp)
e^{i\vp\cdot\vec x} ,
\nonumber \\
\ch_A(t;\vp)
&=&b_A(t;\vp)u_L(\vp)
+(\cmat d)_A(t;\vp)u_R(\vp)
\nonumber \\
&&+(\cmat b)_A^*(t;-\vp)v_R(-\vp)
+d_A^*(t;-\vp)v_L(-\vp) .
\nonumber \\
\label{nup}\eea
This is chosen to satisfy explicitly 
the charge-conjugation condition $\ch^C_A=\cmat_{AB}\ch_B$.
The spinor basis
$\{u_L(\vp), u_R(\vp),v_R(-\vp), v_L(-\vp)\}$
obeys the usual relations for massless fermions,
with $v_{R,L}(\vp)=C\bar u^T_{L,R}(\vp)$.
It has eigenvalues of the helicity operator
$\ga_5\ga^0\vec\ga\cdot\vp/|\vp|$
given by $\{-,+,-,+\}$
and eigenvalues of the chirality operator $\ga_5$
given by $\{-,+,+,-\}$.
For simplicity, we normalize with
$u_\al^\dag u_\be = v_\al^\dag v_\be =\de_{\al\be}$
for $\al,\be = L,R$.
The definition \rf{nup} implies that the amplitudes
$b_{e, \mu, \ta}$
may be approximately identified with active neutrinos
and $d_{e, \mu, \ta}$ with active antineutrinos.
The remaining amplitudes
$b_{e^C,\mu^C,\ta^C}$ and $d_{e^C,\mu^C,\ta^C}$
cover the space of sterile right-handed neutrinos,
but a simple identification with flavor neutrinos and antineutrinos
would be inappropriate in view of their large mass.

In the mass-diagonal Majorana basis,
we restrict attention to the propagating states 
consisting of the light neutrinos.
Taking the hamiltonian in this basis,
\beq
\cH_{\aa \bb}(\vp)=
\ga^0(\vec\ga\cdot\vp  +m_{(\aa )})
\de_{\aa \bb}
+\de \cH_{\aa \bb}(\vp),
\eeq
and applying it to
$\ch_{\bb}(t;\vp)
=U_{\bb B}\ch_B(t;\vp)$
yields the equations of motion
in terms of the amplitudes $b$ and $d$.
The result takes the form of the matrix equation
\beq
[i\de_{\aa \bb}\prt_0
-H_{\aa \bb}(\vp)]
\left(
\begin{array}{c}
b_\bb(t;\vp) \\
d_\bb(t;\vp) \\
b_\bb^*(t;-\vp) \\
d_\bb^*(t;-\vp)
\end{array}
\right)=0 ,
\label{we}
\eeq
where for convenience we have defined 
$b_\bb=V_{\bb B}b_B$ and
$d_\bb=V_{\bb B}^*d_B$,
and where $H_{\aa \bb}$ 
is the spinor-decomposed form of 
$\cH_{\aa \bb}$.

The propagation of kinematically allowed states 
is completely determined by the amplitudes $b_\aa$ and $d_\aa$.
However, 
for purposes of comparison with experiment
it is convenient to express the result using
the amplitudes associated with active neutrinos,
$b_{e,\mu,\ta}$ and $d_{e,\mu,\ta}$.
The relevant calculation is somewhat lengthy
and is deferred to Appendix \ref{hcalc}.
It assumes that the submatrix $V_{\aa a}$ is unitary,
and it neglects terms that enter 
as small masses $m_{(\aa)}$ multiplied by 
coefficients for Lorentz violation,
since these are typically suppressed.
The calculation reveals that the time evolution 
of the active-neutrino amplitudes
is given by the equation
\beq
\left(
\begin{array}{c}
b_a(t;\vp)\\
d_a(t;\vp)
\end{array}
\right)
=\exp(-i\heff t)_{ab}
\left(
\begin{array}{c}
b_b(0;\vp)\\
d_b(0;\vp)
\end{array}
\right) ,
\label{Ut}
\eeq
where $\heff$ is the effective hamiltonian
describing flavor neutrino propagation.
To leading order,
it is given by
\begin{widetext}
\bea
(\heff)_{ab}&=&
|\vp|\de_{ab}
\left(\begin{array}{cc}
1 & 0 \\
0 & 1
\end{array}\right)
+\fr{1}{2|\vp|}
\left(\begin{array}{cc}
(\mt)_{ab} & 0 \\
0& (\mt)^*_{ab}
\end{array}\right)
\nonumber \\
&&\quad+\fr{1}{|\vp|}
\left(\begin{array}{cc}
[(a_L)^\mu p_\mu-(c_L)^\mn p_\mu p_\nu]_{ab} &
-i\sqrt{2} p_\mu (\ep_+)_\nu
[(g^{\mn\si}p_\si-H^\mn)\cmat]_{ab} \\
i\sqrt{2} p_\mu (\ep_+)^*_\nu
[(g^{\mn\si}p_\si+H^\mn)\cmat]^*_{ab} &
[-(a_L)^\mu p_\mu-(c_L)^\mn p_\mu p_\nu]^*_{ab}
\end{array}\right) ,
\label{heff}
\eea
\end{widetext}
where we have defined
$(c_L)^\mn_{ab}\equiv(c+d)^\mn_{ab}$ and
$(a_L)^\mu_{ab}\equiv(a+b)^\mu_{ab}$
for reasons explained below.
The approximate four momentum $p_\mu$ 
may be taken as $p_\mu=(|\vp|;-\vp)$
at leading order.
The Lorentz-conserving mass term 
results from the usual seesaw mechanism 
with $\mt\equiv\ml \ml^\dag$,
where $\ml$ is the light-mass matrix
$\ml= L - DR^{-1}D^T$.
The complex vector $(\ep_+)_\mu$ satisfies
the conditions
\bea
p^\mu(\ep_+)^\nu-p^\nu(\ep_+)^\mu
&=&i\ep^{\mn\rh\si}
p_\rh(\ep_+)_\si ,\nonumber \\
(\ep_+)^\nu(\ep_+)^*_\nu&=&-1 .
\label{ep}
\eea
A suitable choice is
$(\ep_+)^\nu=\frac{1}{\sqrt{2}}(0;\hat\ep_1+i\hat\ep_2)$,
where $\hat\ep_1$, $\hat\ep_2$ are real
and $\{ \vp/|\vp|, \hat\ep_1, \hat\ep_2 \}$
form a right-handed orthonormal triad.
Note that $(\ep_+)^\nu$ and
$(\ep_-)^\nu\equiv(\ep_+)^{\nu*}$
is analogous to the usual photon helicity basis.
The appearance of these vectors
reflects the near-definite helicity of active neutrinos.
The vectors $\hat\ep_1$ and $\hat\ep_2$
can be arbitrarily set by rotations
or equivalently by multiplying $(\ep_+)^\nu$ by a phase,
which turns out to be equivalent to changing the relative phase 
between the basis spinors $u_L$ and $u_R$.

Only the diagonal kinetic term in $\heff$
arises in the minimal SM.
The term involving $(\mt)_{ab}$ 
encompasses the usual massive-neutrino case 
without sterile neutrinos.
The leading-order Lorentz-violating contributions 
to neutrino-neutrino mixing are controlled by 
the coefficient combinations 
$(a+b)^\mu_{ab}$ and $(c+d)^\mn_{ab}$.
These combinations conserve 
the usual SU(3)$\times$SU(2)$\times$U(1) gauge symmetry 
and correspond to the coefficients
$(a_L)^\mu_{ab}$ and $(c_L)^\mn_{ab}$
in the minimal SME.
Note that the orthogonal combinations
$(a-b)^\mu_{ab}$ and $(c-d)^\mn_{ab}$
also conserve the usual gauge symmetry,
but they correspond to self-couplings of right-handed neutrinos 
and are therefore irrelevant for leading-order processes
involving active neutrinos.
The remaining coefficients,
$(g^{\mn\si}\cmat)_{ab}$ and $(H^\mn\cmat)_{ab}$,
appear in $\heff$ through Majorana-like couplings 
that violate SU(3)$\times$SU(2)$\times$U(1) gauge invariance
and lepton-number conservation.
They generate Lorentz-violating neutrino-antineutrino mixing.

Some combinations of coefficients may be unobservable,
either due to symmetries 
or because they can be removed through field redefinitions
\cite{ck,kle,cm,bek}.
For example, 
the trace component $\et_\mn(c_L)^\mn$ is Lorentz invariant
and can be absorbed into the usual kinetic term,
so it may be assumed zero for convenience.
In fact,
even if this combination is initially nonzero,
it remains absent from the leading-order effective hamiltonian
because the trace of $p_\mu p_\nu$ vanishes.
Other examples of unobservable coefficients include 
certain combinations of $g^{\mn\si}$ and $H^\mn$.
The antisymmetry properties
$g^{\mn\si}=-g^{\nu\mu\si}$, $H^\mn=-H^{\nu\mu}$
and the properties of $(\ep_+)_\nu$ can be combined 
to prove that the physically significant combinations 
of $g^{\mn\si}$ and $H^\mn$ are given by the relations
\bea
p_\mu (\ep_+)_\nu g^{\mn\si}
=|\vp|(\ep_+)_\nu \gt^{\nu\si},
\nonumber \\
p_\mu (\ep_+)_\nu H^\mn
=|\vp|(\ep_+)_\nu \Ht^\nu, 
\eea
where we have defined 
\bea
\gt^{\nu\si}\equiv
g^{0\nu\si}
+\frac{i}{2}{\ep^{0\nu}}_{\ga\rh}g^{\ga\rh\si} ,
\nonumber \\
\Ht^\nu\equiv
H^{0\nu}
+\frac{i}{2}{\ep^{0\nu}}_{\ga\rh}H^{\ga\rh} .
\label{gtHt}
\eea
Only these combinations appear in $\heff$
and are relevant to neutrino oscillations.

In deriving Eq.\ \rf{heff},
we have focused on operators of renormalizable dimension,
which involve linear derivatives in the equations of motion
and a single power of momentum in the hamiltonian.
Operators of nonrenormalizable mass dimension $n>4$ are also of 
potential importance
\cite{kpo,kle}.
They appear as higher-derivative terms in the action,
along with corresponding complications in the equations of motion
and in the construction of the hamiltonian. 
An operator of dimension $n$ is associated 
with a term in the action involving $d=n-3$ derivatives,
and the associated terms in the effective hamiltonian
involve $d$ powers of the momentum. 
The corresponding coefficient for Lorentz violation 
carries $d+2$ or fewer Lorentz indices,
depending on the spinor structure of the coupling
and the number of momentum contractions occurring.
For the case $n>4$,
we generically denote the coefficients by $(k_d)^{\la\ldots}$.
These coefficients have mass dimension $1-d$.
Note that,
depending on the theory considered,
the mechanism for Lorentz and CPT violation
can cause them to be suppressed
by $d$-dependent powers of the Planck scale
\cite{kpo,kle}.
Some effects of operators with $d=2$ have been considered
in the context of quantum gravity in Ref.\ \cite{bef}. 

The mixing described by Eq.\ \rf{heff}
or its generalization to operators of dimension $n>4$
can be strongly energy dependent.
For example,
any nonzero mass-squared differences 
dominate the hamiltonian at some low-energy scale.
However,
while mass effects decrease with energy,
Lorentz-violating effects 
involving operators of renormalizable dimension 
remain constant or grow linearly with energy $E$
and so always dominate at high energies.
For instance, 
the contributions from a mass of $0.1$ eV
and a dimensionless coefficient of $10^{-17}$
are roughly comparable at an energy
determined by $E^2\sim (0.1$ eV$)^2/(10^{-17})$,
or $E\sim 30$ MeV.
Below this energy the mass term dominates, 
while above it the Lorentz-violating term does.
Similarly, 
a dimension-one coefficient of $10^{-15}$ GeV
has a transition energy $E\sim10$ keV.
More generally,
effects controlled by the coefficients $(k_d)^{\la\ldots}$
for Lorentz violation involving operators of dimension $n=d+3$
grow as $E^d$. 

Although the perturbative diagonalization leading to Eq.\ \rf{heff} 
is valid for dimensionless coefficients much smaller than one
and for energies much greater than any masses 
or coefficients of dimension one,
at sufficiently high energies
issues of stability and causality 
may require the inclusion of Lorentz-violating terms 
of nonrenormalizable dimension in the theory. 
In the context of the single-fermion QED extension,
for example,
a dimensionless $c^{00}$ coefficient
can lead to issues with causality and stability 
at energies $\sim m_{\rm fermion}/\sqrt{c^{00}}$
unless the effects of operators of nonrenormalizable dimension 
are incorporated
\cite{kle}.
A complete resolution of this issue would be of interest
but lies beyond our present scope.
It is likely to depend on the underlying mechanisms
leading to mass and Lorentz violation,
and it may be complicated further by the presence 
of multiple generations and the sterile neutrino sector. 
We limit our remarks here to noting that  
the values of the coefficients for Lorentz violation
considered in all the models in this work
are sufficiently small that
issues of stability and causality can be arranged 
to appear only beyond experimentally relevant energies.
In any case, 
the renormalizable sector 
provides a solid foundation for the basic treatment
of Lorentz and CPT violation in neutrinos. 

\subsection{Neutrinos in matter}\label{matter}

In many situations,
neutrinos traverse a significant volume of ordinary matter 
before detection.
The resulting forward scattering with electrons, protons, and neutrons 
can have dramatic consequences on neutrino oscillations
\cite{kuop}.
These matter interactions can readily be incorporated
into our general formalism.
Since the effective lagrangian in normal matter is given by 
$\De\cl_{\rm matter}=
-\sqrt{2}G_Fn_e\nub_e\ga^0P_L\nu_e
+(G_Fn_n/\sqrt{2})\nub_a\ga^0P_L\nu_a$,
matter effects are equivalent to contributions
from CPT-odd coefficients
\bea
(a_{L,\rm eff})^0_{ee}&=&G_F(2n_e-n_n)/\sqrt{2} ,
\nonumber \\
(a_{L,\rm eff})^0_{\mu\mu}&=&(a_{L,\rm eff})^0_{\ta\ta}=
-G_Fn_n/\sqrt{2} ,
\eea
where $n_e$ and $n_n$ are the number
densities of electrons and neutrons.
Adding these terms to the effective hamiltonian \rf{heff}
therefore incorporates the effects of matter.

For some of the analyses of Lorentz violation below,
it is useful to review the treatment of matter effects
in solar and atmospheric neutrinos. 
Consider first solar neutrinos.
These are produced in several processes
that generate distinct, well-understood $\nu_e$ spectra.
The most notable are the pp spectrum 
with a maximum energy of about 0.4 MeV,
and the $^8$B spectrum
with a maximum of about 16 MeV
\cite{ssm}.
For $\nu_a\mix\nu_b$ mixing scenarios,
the contribution from $n_n$ is the same for all species
and therefore can be ignored.
However,  
$n_n$ may be important for $\nu_a\mix\nub_b$ mixing,
such as that generated by the coefficients
$(g^{\mn\si}\cmat)_{ab}$ and $(H^\mn\cmat)_{ab}$ in $\heff$.
An analytic approximation to the electron number density 
inside the Sun is given by \cite{ssm}
$n_e/N_A=245e^{-10.54R/R_\odot}$.
It is useful to define $n_s=n_e-\half n_n$,
a combination that often appears in sterile-neutrino searches.
This number density has a similar approximation,
$n_s/N_A=223e^{-10.54R/R_\odot}$.
The two linearly independent combinations can therefore be taken as
$G_Fn_e\simeq 1.32\times10^{-20}e^{-10.54R/R_\odot}$ GeV
and
$G_Fn_s\simeq 1.20\times10^{-20}e^{-10.54R/R_\odot}$ GeV,
corresponding to a neutron contribution of
$G_Fn_n=2G_F(n_e-n_s)\simeq 0.24\times10^{-20}e^{-10.54R/R_\odot}$ GeV
to the effective hamiltonian.
These quantities set the scale for matter effects in the Sun.

Next, consider the detection of atmospheric neutrinos.
Upward-going neutrinos pass through the Earth 
and therefore experience higher matter potentials
than the downward-going neutrinos,
which traverse the less dense atmosphere and a small amount
of bedrock on their way to the detector.
A crude estimate of the matter potential in this case 
can be obtained by assuming that the Earth
consists of roughly equal numbers of 
protons, neutrons, and electrons.
Using the average number density then
yields the approximate value
$G_Fn_e\simeq G_Fn_n\simeq 1.5\times10^{-22}$ GeV.
This produces a matter potential similar to that 
from the Sun at $R/R_\odot\sim 2/5$.

Overall,
the contribution to $\heff$ from matter ranges from about
$10^{-20}$ GeV to $10^{-25}$ GeV.
This means that matter effects must be incorporated
when the contributions from mass or Lorentz violation 
lie near these values.
This range is comparable to the scale 
of coefficients for Lorentz violation 
that originate as suppressed effects from the Planck scale. 
Note also that most terrestrial experiments
involve neutrinos that traverse at least some amount of bedrock 
or other shielding materials,
which can result in substantially different conventional 
or Lorentz-violating dynamics 
relative to the vacuum-oscillation case
\cite{mattercpt}.

\subsection{Neutrino oscillations}\label{nuosc}

The analysis of neutrino mixing proceeds along the usual lines.
The effective hamiltonian can be diagonalized 
with a $6\times 6$ unitary matrix $\Ueff$:
\beq
\heff =  \Ueff^\dag E_{\rm eff} \Ueff ,
\label{diaheff}
\eeq
where $E_{\rm eff}$ is a $6\times 6$ diagonal matrix.
In contrast to the Lorentz-covariant case,
where mixing without sterile neutrinos involves only three propagating states,
here mixing without sterile neutrinos may occur with six states.
This means that there can be 
up to five energy-dependent eigenvalue differences
for Lorentz-violating mixing,
resulting in five independent oscillation lengths
instead of the usual two.

Denoting the six propagation states 
by the amplitudes $B_J(t;\vp)$
with $J=1,\ldots ,6$, 
we can write 
$B_J(t;\vp)=\tu_{Ja}b_a(t;\vp)+\bu_{Ja}d_a(t;\vp)$,
where we have split $\Ueff$ into $6\times 3$ matrices
$\Ueff=(\tu, \bu)$.
The time evolution operator may then be written as
\bea
S_{ab}(t)
&=& (\Ueff^\dag e^{-iE_{\rm eff}t} \Ueff)_{ab}
\nonumber \\
&=& \left(\begin{array}{cc}
S_{\nu_a\nu_b}(t) & S_{\nu_a\nub_b}(t)\\
S_{\nub_a\nu_b}(t) & S_{\nub_a\nub_b}(t)
\end{array}\right)
\nonumber \\
&=&\sum_{J} e^{-itE_{(J)}}
\left(\begin{array}{cc}
\tu^*_{Ja}\tu_{Jb} & \tu^*_{Ja}\bu_{Jb} \\
\bu^*_{Ja}\tu_{Jb} & \bu^*_{Ja}\bu_{Jb}
\end{array}\right) ,
\label{Ut2}\eea
where $E_{(J)}$ are the diagonal values of $E_{\rm eff}$.

The probabilities for a neutrino of type $b$
oscillating into a neutrino or antineutrino 
of type $a$ in time $t$ are therefore 
$P_{\nu_b\to\nu_a}(t)=|S_{\nu_a\nu_b}(t)|^2$
or 
$P_{\nu_b\to\nub_a}(t)=|S_{\nub_a\nu_b}(t)|^2$,
respectively.
Similarly, 
for antineutrinos we have
$P_{\nub_b\to\nu_a}(t)=|S_{\nu_a\nub_b}(t)|^2$
or 
$P_{\nub_b\to\nub_a}(t)=|S_{\nub_a\nub_b}(t)|^2$.
In terms of the matrices
$\tu$ and $\bu$,
the probabilities are
\begin{widetext}
\bse\bea
P_{\nu_b\to\nu_a}(t)&=&\de_{ab}
-4\sum_{J>K}\Re(\tu^*_{Ja}\tu_{Jb}\tu_{Ka}\tu^*_{Kb})
\sin^2\fr{\De_{JK}t}2
+2\sum_{J>K}\Im(\tu^*_{Ja}\tu_{Jb}\tu_{Ka}\tu^*_{Kb})
\sin\De_{JK}t\ ,
\label{pnn}\\
P_{\nub_b\to\nub_a}(t)&=&\de_{ab}
-4\sum_{J>K}\Re(\bu^*_{Ja}\bu_{Jb}\bu_{Ka}\bu^*_{Kb})
\sin^2\fr{\De_{JK}t}2
+2\sum_{J>K}\Im(\bu^*_{Ja}\bu_{Jb}\bu_{Ka}\bu^*_{Kb})
\sin\De_{JK}t\ ,
\label{pbb}\\
P_{\nu_b\to\nub_a}(t)&=&
-4\sum_{J>K}\Re(\bu^*_{Ja}\tu_{Jb}\bu_{Ka}\tu^*_{Kb})
\sin^2\fr{\De_{JK}t}2
+2\sum_{J>K}\Im(\bu^*_{Ja}\tu_{Jb}\bu_{Ka}\tu^*_{Kb})
\sin\De_{JK}t\ ,
\label{pnb}\\
P_{\nub_b\to\nu_a}(t)&=&
-4\sum_{J>K}\Re(\tu^*_{Ja}\bu_{Jb}\tu_{Ka}\bu^*_{Kb})
\sin^2\fr{\De_{JK}t}2
+2\sum_{J>K}\Im(\tu^*_{Ja}\bu_{Jb}\tu_{Ka}\bu^*_{Kb})
\sin\De_{JK}t\ ,
\label{pbn}
\eea\label{probs}\ese
\end{widetext}
where the effective-energy difference
is denoted by $\De_{JK}=E_{(J)}-E_{(K)}$.

\subsection{CPT properties}\label{cptprop}

With a conveniently chosen phase,
CPT may be implemented by the transformation
\beq
\left(\begin{array}{c}
b^{\rm CPT}_a(t;\vp)\\
d^{\rm CPT}_a(t;\vp)
\end{array}\right)
=i\left(\begin{array}{c}
-d^*_a(-t;\vp)\\
b^*_a(-t;\vp)
\end{array}\right)
\equiv \si^2
\left(\begin{array}{c}
b^*_a(-t;\vp)\\
d^*_a(-t;\vp)
\end{array}\right) .
\label{cpt}
\eeq
This yields precisely the expected
result when applied to $\heff$:
the CPT-conjugate hamiltonian
$h^{\rm CPT}_{\rm eff}=\si^2\heff^*\si^2$
can be obtained from Eq.\ \rf{heff}
by changing the sign of the CPT-odd
$a_L$ and $g$ coefficients.
Then,
$h^{\rm CPT}_{\rm eff}=\heff$ when $a_L$ and $g$ vanish,
as expected.
A notable feature here is that independent mass matrices
for neutrinos and antineutrinos
cannot be generated as has been proposed 
\cite{mmb}.
Greenberg has recently proved that this result
is general 
\cite{owg}.

Under CPT, 
the transition amplitudes transform as
\bse\bea
S_{\nu_a\nu_b}(t)
&\stackrel{\rm CPT}{\longleftrightarrow}&
S^*_{\nub_a\nub_b}(-t),
\label{Scpt1} \\
S_{\nub_a\nu_b}(t)
&\stackrel{\rm CPT}{\longleftrightarrow}&
-S^*_{\nu_a\nub_b}(-t) .
\label{Scpt2}
\eea\label{Scpt}\ese
These relations become equalities if CPT holds.
The first relation then yields the usual result,
\bse
\beq
\mbox{\rm CPT invariance }
\implies
P_{\nu_b\to\nu_a}(t) = P_{\nub_a\to\nub_b}(t) .
\label{cpt1}
\eeq
This property has long been understood
and has been identified as a potential test of CPT invariance
\cite{fc2}.
However, 
the negation of terms in this result
produces a statement that may be false in general
because CPT violation need not imply 
$P_{\nu_b\to\nu_a}(t) \neq 
P_{\nub_a\to\nub_b}(t)$.
Examples of models that violate CPT but nonetheless
satisfy Eq.\ \rf{cpt1} are given in Sec.\ \ref{models}.

The above property addresses the relationship between 
$\nu\mix\nu$ and $\nub\mix\nub$ mixing.
There is also an analogous property
associated with $\nu\mix\nub$ mixing.
Thus,
for CPT invariance,
relation \rf{Scpt2} yields the additional result:
\beq
\mbox{\rm CPT invariance }
\implies
P_{\nu_b\rightleftarrows\nub_a}(t) = 
P_{\nu_a\rightleftarrows\nub_b}(t).
\label{cpt2}
\eeq
\ese
This property may also provide opportunities 
to test for Lorentz and CPT invariance.
Note,
however,
that negation of its terms produces a statement that may be false in general,
as in the previous case.

Finally,
we emphasize that the presence of CPT violation increases the 
number of independent oscillation lengths 
without the addition of sterile neutrinos.
In the general case,
nonzero coefficients for CPT violation 
in the effective hamiltonian \rf{heff}
can generate up to six independent propagating states,
rather than the usual three.

\subsection{Reference frames} \label{frames}

The presence of Lorentz violation makes 
it necessary to specify the frame 
in which experimental results are reported.
Coordinate invariance of the physics,
in particular observer Lorentz invariance
\cite{ck},
ensures that the analysis and measurements of an experiment
can be performed in any frame of reference.
However,
it is convenient to have a standard
set of frames to facilitate comparisons of different experiments.
In the literature,
measurements are conventionally expressed 
in terms of coefficients for Lorentz violation
defined in a Sun-centered celestial equatorial frame 
with coordinates $(T,X,Y,Z)$
\cite{fn1}.
For our present purposes, 
it suffices to identify the $Z$ direction 
as lying along the Earth's rotational axis and the
$X$ direction as pointing towards the vernal equinox.
The coefficients for Lorentz violation
in any other inertial frame can be related 
to the standard set in the Sun-centered frame 
by an observer Lorentz transformation.
In general, 
this transformation includes both rotations and boosts,
but boost effects are frequently neglected
because they introduce only terms suppressed by
the velocity $\be$ between frames,
which is typically $\lsim 10^{-4}$.
Recently, 
studies of some $\be$-suppressed terms have been performed
in the context of high-precision clock-comparison experiments
\cite{spaceexpt,cane}
and resonant cavities
\cite{cavexpt,km}.

The existence of orientation-dependent effects 
makes it useful to define a standard parametrization 
for the direction of neutrino propagation $\hat p$ 
and the corresponding $\hat\ep_1$, $\hat\ep_2$ vectors
in the Sun-centered frame.
A suitable set of unit vectors is given by
\bea
\hat p&=&(\sin\Th\cos\Ph,\sin\Th\sin\Ph,\cos\Th) ,\nonumber\\
\hat\ep_1&=&(\cos\Th\cos\Ph,\cos\Th\sin\Ph,-\sin\Th) , \nonumber\\
\hat\ep_2&=&(-\sin\Ph,\cos\Ph,0) ,
\label{vectors}
\eea
where $\Th$ and $\Ph$ 
are the celestial colatitude and longitude of propagation,
respectively.
We remark that these quantities are related 
to the right ascension $r$ and declination $d$ 
of the source as viewed from the detector
by $\Th=90^\circ+d$ and $\Ph=180^\circ+r$.

In the remainder of this subsection,
we provide some technical comments about the frame-dependence of 
our choice of spinor basis in Sec.\ \ref{phenom}.
This basis is normally associated with massless fermions,
so the presence of mass or Lorentz violation 
means that even with a covariant normalization
the corresponding amplitudes 
are no longer scalar functions 
under observer Lorentz transformations
and hence are frame dependent.
However,
our basis suffices for perturbative calculations
in which the physically significant states are affected 
only by masses and coefficients for Lorentz violation 
that are small relative to $|\vp|$,
while the complexity of the general Lorentz-violating case 
makes the decomposition into a covariant basis impractical.
Moreover, 
despite the frame-dependent nature of the calculation,
the probabilities \rf{probs} are frame independent at leading order.
In the usual case,
frame independence follows from the Lorentz-vector nature
of the exact 4-momenta $(E_{(J)};\vp)$,
which implies the products
$E_{(J)}t-\vp\cdot\vec x$
are Lorentz scalars, 
and from the constancy and frame-independence 
of the mixing matrix $\Ueff$.
It turns out that a version of these properties 
holds in the present case,
as we show next.

First,
we observe that
the elements of the $6\times 6$ matrix 
$|\vp|(\heff-|\vp|)$
are scalars under observer Lorentz transformations
at leading order in small quantities.
Next, 
note that the matrix $\Ueff$ diagonalizes
$|\vp|(\heff-|\vp|)$,
so its elements can be chosen to be observer Lorentz scalars as well.
In turn, 
this means that the diagonal elements
$|\vp|(E_{(J)}-|\vp|)$
are also observer Lorentz scalars,
since they are functions of the elements of
$|\vp|(\heff-|\vp|)$.
From this result,
it follows explicitly that
the neutrino dispersion relations
$E_{(J)}^2-\vp^{\, 2}$
are observer Lorentz scalars at leading order,
since
\bea
E_{(J)}^2-\vp^{\, 2}
&=&(E_{(J)}+|\vp|)(E_{(J)}-|\vp|)
\nonumber \\
&\simeq&2|\vp|(E_{(J)}-|\vp|) .
\label{disp}
\eea
The 4-momentum is therefore a vector 
under observer Lorentz transformations to leading order,
as desired.
Combining this property with the scalar character of $\Ueff$
implies that the leading-order transition amplitudes 
and probabilities \rf{probs} 
are covariant under observer Lorentz transformations,
as claimed.


\section{Sensitivities}\label{sens}

\subsection{Existing constraints}

To date, 
there is no compelling experimental evidence 
for nonzero coefficients for Lorentz violation
in any sector.
Theoretical predictions of the size of the effects
depend on the underlying model.
However,
the natural scale for a fundamental theory is the Planck mass $m_P$,
which is about 17 orders of magnitude greater than the
electroweak scale $m_W$ relevant to the SM 
and roughly 30 orders of magnitude greater 
than the scale of neutrino masses, 
if they exist. 
It is plausible that any observable 
Lorentz- and CPT-violating effects 
are suppressed by one or more powers of the dimensionless ratio 
$r = m/m_P \lsim 10^{-17}$,
where $m$ is the relevant low-energy scale
and $m_P$ is the Planck mass  
\cite{kp}.
In contrast,
the scale of observed neutrino oscillations is $\lsim 0.1$ eV,
which enters as a squared mass $\De m^2 \lsim 10^{-20}$ GeV.
At physically relevant energies,
$10^{-4}$ GeV $< E < 10^3$ GeV,
the oscillation physics is determined
by the dimensionless ratio $r_\nu = \De m^2/E^2$. 
Remarkably,
the two dimensionless ratios $r$ and $r_\nu$ have a similar range,
so the natural size of Lorentz- and CPT-violating effects 
may be comparable to the natural size of neutrino-oscillation effects.

Certain experiments in the fermion and photon sectors 
have achieved sensitivities corresponding to
dimensionless suppressions of roughly $10^{-30}$.
Since the coefficients for Lorentz violation 
in the various sectors can be related either directly 
through symmetries or indirectly through radiative corrections,
it might seem that existing experimental constraints 
severely restrict the possibilities for Lorentz violation in neutrinos.
In fact,
this expectation is incorrect,
as we discuss next.

In the context of $\heff$,
the relevant coefficients are
$(a_L)^\mu_{ab}$ and $(c_L)^\mn_{ab}$,
since these appear directly in the charged-fermion sector of the SME.
A decomposition of the multi-flavor QED limit 
of the charged-lepton sector 
can be performed in analogy with Eq.\ \rf{GaM}.
It produces the identification
\bea
a^\mu_{ab}&=&\half(a_L+a_R)^\mu_{ab} ,\nonumber\\
b^\mu_{ab}&=&\half(a_L-a_R)^\mu_{ab} ,\nonumber\\
c^\mn_{ab}&=&\half(c_L+c_R)^\mn_{ab} ,\nonumber\\
d^\mn_{ab}&=&\half(c_L-c_R)^\mn_{ab} ,
\eea
where $(c_R)^\mn_{ab}$ and $(a_R)^\mu_{ab}$
are coefficients in the SME that couple to right-handed leptons
and therefore leave unaffected the active neutrinos at tree level.
On this basis,
it might naively appear that the charged sector is sensitive 
to more combinations of coefficients for Lorentz violation 
than the neutrino sector.
However, 
the mass hierarchy of the charged leptons
$e$, $\mu$, $\ta$ 
implies that only coefficients that are diagonal in flavor space 
appear in leading-order perturbative calculations.
As a result, 
$e$, $\mu$, $\ta$ effectively decouple,
resulting in three independent copies 
of the fermion sector 
in the Lorentz- and CPT-violating QED extension.
This implies that 
unsuppressed sensitivity to Lorentz violation
in the charged-lepton sector 
involves only flavor-diagonal components.
Moreover, 
the decoupling also implies that 
certain coefficients 
such as $a^\mu_{ee}$, $a^\mu_{\mu\mu}$, $a^\mu_{\ta\ta}$ 
are physically unobservable, 
further reducing the total number of coefficients 
affecting charged leptons.
Taken together, 
these factors ensure that the CPT-odd sectors
of charged leptons and neutrinos 
are completely independent at tree level.
Similar arguments apply to parts of the CPT-even sector as well.
We therefore conclude that
neutrinos are sensitive to a greater number 
of coefficients for Lorentz violation than the charged leptons,
and at tree level most of these coefficients are independent 
from those accessible with $e$, $\mu$ or $\ta$ leptons.

Particularly stringent constraints
exist on some components of the charged-lepton coefficients
$b^\mu_{ee}$ and $b^\mu_{\mu\mu}$.
Although these are linearly independent
of neutrino-sector coefficients at tree level,
it is natural to ask whether
radiative corrections to these components
can be used to constrain possible neutrino effects.
As an example,
Ref.\ \cite{fc5} explores the possibility that
eV-size effects in heavy sterile neutrinos 
could evade the constraints in the charged-lepton sector,
finding that within a standard seesaw mechanism
the existence of large $b^\mu$-type coefficients 
for sterile neutrinos tends to produce $b^\mu$ coefficients 
in the charged-lepton sector that conflict with observation.
In this work, 
we neglect seesaw-induced coefficients 
because they are suppressed by the large-mass scale.
However,
it is of interest to ask whether radiative corrections 
alter the tree-level independence of the charged- and
neutral-lepton sectors.

For simplicity,
we restrict our discussion 
to the relevant $a^\mu$ and $b^\mu$ coefficients, 
although related remarks 
apply also to $c^\mn$ and $d^\mn$ coefficients.
The leading-order radiative corrections are linear
in the coefficients for Lorentz violation.
However,
loops involving weak-interactions are heavily suppressed
by additional factors at the relevant energies,
while strong interactions play no role. 
We can therefore restrict attention 
to the QED extension.
In this case,
general properties of the coefficients for Lorentz violation
under the discrete symmetries C, P, and T
imply that corrections to $b^\mu$ coefficients 
involve only other $b^\mu$ type coefficients \cite{klap}.
As a result,
although the constraints from charged-lepton experiments 
may restrict $b^\mu$ in the neutrino sector of the SME, 
the $a^\mu$ coefficients are unaffected
and so $a_L$ is unconstrained.
Thus,
the independence of the charged- and neutral-lepton sectors 
remains valid for radiative corrections.

\subsection{General features}\label{features}

In the presence of Lorentz and CPT violation,
a wide range of unconventional neutrino behaviors can occur.
These include unusual energy dependence,
direction-dependent effects, 
and neutrino-antineutrino mixing.
Specific examples of these behaviors
are illustrated in the examples presented 
in Sec.\ \ref{models}.
Here,
we focus on some general features 
of experimental sensitivities to Lorentz- and CPT-violating effects. 
Some of these have been discussed in the context of
the minimal SME in our earlier work
\cite{nu},
but the present discussion holds for the full theory \rf{heff}
and generically for operators of nonrenormalizable dimension.

Figure 1 shows an estimate of the coverage 
in baseline distance $L$ versus energy $E$
of the currently published neutrino-oscillation data.
Included in the evidence for oscillations are observations of
solar neutrinos by 
Cl- and Ga-based experiments
\cite{gallex,gno,sage,homestake},
Super Kamiokande (SK) \cite{sksol},
and SNO \cite{sno}; 
and of atmospheric neutrinos by
SK \cite{sk},
reactor-based KamLAND
\cite{kamland},
and accelerator-based
LSND \cite{lsnd} 
and K2K \cite{k2k}.
Null results include the reactor experiments
Bugey \cite{bugey},
CHOOZ \cite{chooz},
G\"osgen \cite{gosgen},
Palo Verde \cite{pv},
and various accelerator-based short-baseline experiments including,
for example, 
the high-energy experiments 
BNL-E776 \cite{e776},
CCFR \cite{ccfr},
CHORUS \cite{chorus},
NOMAD \cite{nomad1,nomad2},
NuTeV \cite{nutev},
and the low-energy
KARMEN \cite{karmen}.
A number of new accelerator-based experiments
are likely to produce interesting results in the near future.
These include the short-baseline 
($L\simeq 500$ m, $E\simeq 1$ GeV)
MiniBooNE experiment 
\cite{miniboone}
designed to test the LSND anomaly,
and the long-baseline
($L\simeq 700$ km, $E\simeq 1$ GeV)
ICARUS \cite{icarus},
MINOS \cite{minos},
and OPERA \cite{opera}
experiments,
which are planned to test the atmospheric-oscillation hypothesis.
Also shown on the figure are the approximate effective regions
associated with the matter potentials for the Sun and the Earth.

The unusual energy dependence
can be viewed as a consequence of the dimensionality
of the coefficients for Lorentz violation.
The standard scenario for neutrino oscillations
involves mass-squared differences $\De m^2$
that combine with the baseline distance $L$ and the neutrino energy $E$
to yield the physically relevant dimensionless combination
$\De m^2L/E$.
However,
Eq.\ \rf{heff} shows that Lorentz-violating oscillations 
generated by the dimension-one coefficients
$a^\mu$, $b^\mu$, $H^\mn$
are controlled by the dimensionless combinations
$a^\mu L$, $b^\mu L$, $H^\mn L$,
while those generated by
$c^\mn$, $d^\mn$, $g^{\mn\si}$
are controlled by
$c^\mn LE$, $d^\mn LE$, $g^{\mn\si} LE$.
More generally,
oscillations
generated by a coefficient $(k_d)^{\la\ldots}$
for a Lorentz-violating operator
of nonrenormalizable dimension $n=d+3$
are controlled by $(k_d)^{\la\ldots} LE^d$.

\begin{figure}
\centerline{
\hspace{2pt}\psfig{figure=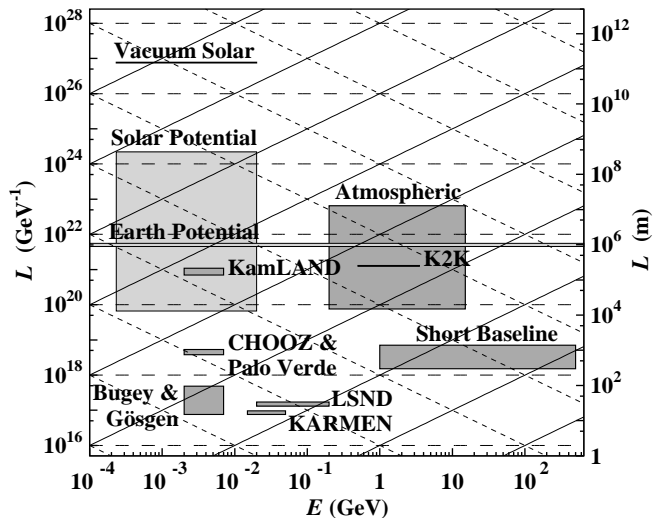,width=1.0\hsize}\hspace{2pt}}
\caption{\label{expts}
Approximate sensitivities of various experiments.
Lines of constant $L/E$ (solid),
$L$ (dashed), and $LE$ (dotted)
are shown,
giving approximate sensitivities
to the quantities $\{m, m_5\}$, $\{a^\mu, b^\mu, H^\mn\}$, and
$\{c^\mn, d^\mn, g^{\mn\si}\}$,
respectively.
Also shown are the approximate effective regions 
for the matter potential in the Sun and Earth.}
\end{figure}

Figure 1 illustrates these various energy dependences.
Lines of constant $L/E$, $L$, and $LE$ are plotted,
bounding approximate regions of experimental sensitivity to 
conventional mass-squared differences, 
dimension-one coefficients,
and dimensionless coefficients,
respectively.
For each nonzero coefficient in $\heff$,
a bounding line on this figure exists
above which the corresponding Lorentz-violating effects 
become of order one.
Given such a line,
any experiments located near or above it 
can be affected by the associated coefficient,
but experiments below it have limited or no sensitivity.
For example,
the region of limiting sensitivity
for a hypothetical dimensionless coefficient
of magnitude $\sim 10^{-18}$
is bounded approximately 
by the dimensionless line satisfying $LE=10^{18}$,
which is the dotted line running just below KamLAND.
Experiments lying above this line,
such as KamLAND, SNO, and SK,
could be sensitive to the effects of this coefficient.
Note that 
approximate regions of experimental sensitivity to 
coefficients $(k_d)^{\la\ldots}$ of dimension $1-d$
could also be identified on the figure.
They would be bounded by lines of constant $LE^d$ with $d>1$,
which have negative-integer slopes. 

Figure 1 also reveals that experiments and data 
allow probes well below the $10^{-17}$ Planck-suppression level.
For instance,
the various null results 
from short-baseline reactor and accelerator experiments
could be reanalysed to yield upper bounds on  
certain coefficients for Lorentz violation.
Thus,
the high-energy experiments CHORUS and NOMAD 
found no evidence of $\nu_{e,\mu}\to\nu_\ta$
at energies $E\sim 100$ GeV 
and at distances $L\sim 10^{18}$ GeV$^{-1}$,
which suggests that reanalyses of these experiments 
would yield interesting new sensitivities 
of roughly $10^{-18}$ GeV to dimension-one coefficients 
and roughly $10^{-20}$ to dimensionless coefficients.
A similar situation holds for low-energy experiments 
such as CHOOZ, Palo Verde, and KARMEN 
in the $\nub_e$ sector.
From Fig.\ \ref{expts} we see that,
relative to CHORUS and NOMAD,
CHOOZ and Palo Verde might be expected 
to have comparable sensitivities 
to dimension-one coefficients
but reduced sensitivity to dimensionless ones,
while KARMEN has comparable sensitivity 
to conventional mass effects.
In each case,
the attainable sensitivities also depend 
on various experiment-dependent factors, 
so individual reanalyses are required to make definitive statements.

Another unusual effect due to Lorentz violation
is direction-dependent neutrino behavior,
a consequence of rotation-symmetry violation.
This has consequences for comparisons of results
between different terrestrial experiments
or for the analysis of experiments involving multiple sources,
since the orientation of the neutrino beam 
or the location of the source relative to the detector
can affect neutrino oscillations.
Rotation-symmetry violation also implies that the
daily rotation of the Earth about its axis induces 
apparent periodic changes of the coefficients for Lorentz violation
in the laboratory,
which would be manifest as temporal variations 
in neutrino oscillations. 
These variations occur at multiples of the sidereal frequency
$\om_\oplus\simeq 2\pi/$(23 h 56 min).
Similarly,
in the presence of rotation-symmetry violation,
neutrinos emitted from the Sun in different directions 
undergo different oscillations,
which may produce observable annual variations 
arising from the change in the location of the detector
as the Earth orbits the Sun.
All these temporal variations with appropriate periodicity 
provide unique signals of Lorentz violation
in neutrino oscillations.
Moreover,
they can also yield interesting sensitivities to certain coefficients.
For instance, 
SK found that the shape of the solar-neutrino flux 
matches the expected value to within about 5\% over the year
\cite{sksol}.
The Sun-Earth distance is $L\sim 10^{27}$ GeV$^{-1}$,
and $LE\sim 10^{25}$ for the SK energy range,
so a reanalysis of the SK data might achieve
impressive sensitivities of $\sim 10^{-28}$ GeV 
to dimension-one coefficients
and $\sim 10^{-26}$ to dimensionless ones,
comparable to the best experimental sensitivities achieved
for other sectors of the SME.

Another interesting feature of Lorentz violation
involves novel resonance effects in neutrino oscillations.
In the conventional case with neutrino masses,
the usual MSW resonances 
\cite{msw}
arise when the local matter environment is such that
neutrino interactions become comparable to mass effects,
thereby drastically changing the character of the hamiltonian.
The presence of Lorentz violation 
can trigger several other types of effects,
including resonances without mass or matter
that involve different coefficients for Lorentz violation,  
resonances involving coefficients for Lorentz violation
and mass terms,  
resonances involving coefficients for Lorentz violation 
and matter effects,  
and various combinations of the above.
The earliest example 
of an explicit vacuum resonance in a two-generation model 
involving a mass term and a single nonzero coefficient $(a_L)^T$ 
for Lorentz and CPT violation is given in Ref.\ \cite{fc2}.
An example of a vacuum resonance 
in a three-generation model 
involving two coefficients $(a_L)^Z$ and $(c_L)^{TT}$ 
for Lorentz and CPT violation 
occurs in the bicycle model of Ref.\ \cite{nu}.
We emphasize that resonances due to Lorentz violation
can occur in the vacuum as well as in matter,
and not only at particular energies 
but also for particular directions of propagation.
Note also that,
even away from the resonance regions,
matter effects may be important
when considering mass terms or coefficients for Lorentz violation 
that have lines of sensitivity near or above 
the Sun- or Earth-potential regions shown in Figure \ref{expts}.

\subsection{The LSND anomaly}\label{lsnd}

In the LSND experiment \cite{lsnd},
copious numbers of neutrinos were
produced from the decay of $\pi^+$ at rest.
This process is dominated by the decay
$\pi^+\to\mu^+\nu_\mu$
followed by $\mu^+\to e^+\nu_e\nub_\mu$.
A small excess in $\nub_e$ was seen,
interpreted as the oscillation
$\nub_\mu\to\nub_e$
with a small probability of about 0.26\%.
This result is difficult to accommodate 
within the context of the conventional global analysis
\cite{pdg},
in which two mass-squared differences
are used to describe solar and atmospheric oscillation data.
The solar data appear consistent 
with a mass-squared difference $\de m^2 \sim 10^{-5}$ eV$^2$,
while the atmospheric data
suggest a second mass-squared difference 
$\De m^2 \sim 10^{-3}$ eV$^2$.
The regions of limiting sensitivity to 
these mass-squared differences 
are shown in Fig.\ \ref{expts}, 
where lines of constant $L/E$ with values
$L/E\sim 10^{23}$ GeV$^{-2}$ and
$L/E\sim 10^{21}$ GeV$^{-2}$
can be seen.
Experiments lying significantly below these lines, 
including LSND, 
should be insensitive to oscillations 
caused by $\de m^2$ and $\De m^2$.
This illustrates the difficulty in explaining the LSND result
within the conventional framework
without introducing additional mass-squared differences.

A resolution of this LSND anomaly
without the introduction of sterile neutrinos 
might emerge from 
the unusual energy dependence,
the directional dependence,
or the neutrino-antineutrino mixing
introduced by Lorentz violation.
For example,
equal numbers of $\nu_\mu$, $\nu_e$, and $\nub_\mu$
are produced in LSND,
so if $\nu_e$ mix with $\nub_e$
then the observed excess in $\nub_e$
may be a result of $\nu_e\mix\nub_e$ mixing
rather than $\nub_\mu\mix\nub_e$ mixing.
We note, 
however,
that if the possible direction dependence is neglected
then Fig.\ \ref{expts} shows that a simple solution based
either on the unusual energy dependence
or on $\nu\mix\nub$ mixing
is likely to be hindered by existing null results
in the $\nub_e$ sector,
from low-energy experiments such as CHOOZ and Palo Verde
or from high-energy experiments such as CHORUS, NOMAD, and NuTeV.
Indeed,
from this figure we see generically that 
to explain the LSND result 
one needs a mass-squared difference of about 
$10^{-19}$ GeV$^2 = 10^{-1}$ eV$^2$,
a dimension-one coefficient of about $10^{-18}$ GeV
or a dimensionless coefficient of about $10^{-17}$.
Note that each of these has consequences for other experiments,
depending on flavor content.
For example, 
the upcoming MiniBooNE experiment is designed to test 
the same oscillation channel 
and will therefore be sensitive to all three possibilities.


\section{Illustrative Models}\label{models}

To illustrate some of the novel behaviors
of neutrino oscillations in the presence of Lorentz and CPT violation,
we next consider a number of simple special cases
of the theory \rf{heff}
with only one or a few nonzero coefficients.
For each case,
some of the ways that the unusual neutrino behaviors
might affect current observations are quantitatively examined.
Also,
we simplify expressions by adopting temporary notation
for the specific nonzero coefficients for Lorentz violation
within each model:
quantities carrying a ring accent, 
such as $\ring{c}$,
denote rotation-symmetric coefficients;
while those with a h\'a\v{c}ek accent,
such as $\check{c}$, 
denote anisotropic coefficients.

\subsection{Rotationally invariant models}\label{rimods}

The rotation-invariant restriction
provides an interesting special limit of the theory \rf{heff}.
While difficult to motivate without knowledge 
of the underlying mechanism leading to Lorentz and CPT violation,
rotation-invariant or so-called `fried-chicken' (FC) models
are attractive because 
rotation symmetry can significantly reduce 
the complexity of calculations,
thereby providing a simple context 
within which to study the unusual neutrino behaviors 
arising from Lorentz violation.

Restricting $\heff$ to FC terms leaves only four matrices,
$(\mt)_{ab}$,
$(a_L)^0_{ab}$,
$(c_L)^{00}_{ab}$,
and
$(c_L)^{jk}_{ab}=\fr13(c_L)^{ll}_{ab}\de^{jk}$.
As described in Sec.\ \ref{phenom},
the trace $(c_L)^{00}_{ab}-(c_L)^{jj}_{ab}$ is
unobservable and may be set to zero,
so only three of these matrices are independent.
Dropping the irrelevant kinetic term
and assuming rotation invariance in the Sun-centered $(T,X,Y,Z)$ frame
for definiteness,
the $6\times6$ effective hamiltonian reduces 
to the block-diagonal form
\bea
(\heff)^{\rm FC}_{ab}&=&{\rm diag}
\big[\big(
\mt/(2E)
+(a_L)^T
-\frac43(c_L)^{TT}E
\big)_{ab}\ ,
\nonumber \\
&&
\quad\big(
\mt/(2E)
-(a_L)^T
-\frac43(c_L)^{TT}E
\big)^*_{ab}\big] .
\label{rim}
\eea
This hamiltonian provides a general FC model
of three active neutrinos.
The generalization to additional light 
or massless sterile neutrinos is straightforward.

With the exception of the original proposal 
for Lorentz violation in neutrinos \cite{ck}
and the recent work in Ref.\ \cite{nu},
which address both rotation-invariant and anisotropic effects 
with and without CPT violation,
existing works on the subject 
\cite{fc1,fc2,fc3,fc4}
involve limited special cases of the general FC model \rf{rim}.
The bulk of the literature
restricts attention to the two-generation special case
and neglects either the $(a_L)^T$ term or the $(c_L)^{TT}$ term.
A plethora of unexplored models and effects exists.

It might seem logical to impose spherical symmetry 
in a special frame 
such as the cosmic microwave background (CMB) frame.
However,
if rotation symmetry is assumed in the CMB frame
then the coefficients in Eq.\ \rf{rim} 
differ from $(a_L)^T$, $(c_L)^{TT}$ 
in the standard Sun-centered frame,
being instead $(a_L)^0$, $(c_L)^{00}$ in the CMB frame.
Relating the latter to the standard Sun-centered frame 
or any other experimentally attainable frame
introduces direction dependence 
due to the motion of our solar system in the CMB frame.
The relevant hamiltonian then also involves spatial
components of the coefficients,
so it differs from Eq.\ \rf{rim} 
and is instead an anisotropic limit of the theory \rf{heff}. 

Although the FC model \rf{rim} is rather limited 
considering the wealth of possible effects 
contained in the full theory \rf{heff},
and although it has little theoretical motivation
other than calculational convenience,
further study of this model is useful 
because it provides a readily workable context 
within which to gain insight 
about possible signals of Lorentz and CPT violation.
This is illustrated in the few simple examples 
discussed in this subsection.

\subsubsection{Example: $(c_L)^{TT}_{ab} \ne 0$}\label{dcmod}

\begin{figure}
\centerline{
\psfig{figure=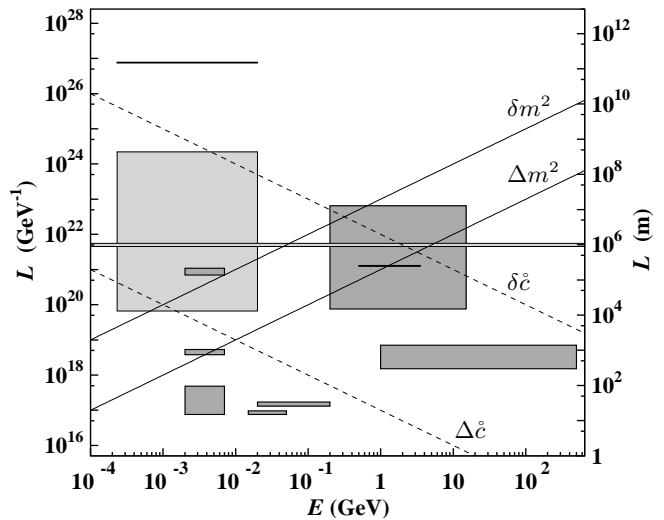,width=1.0\hsize}
\begin{picture}(0,0)(20,0)
\put(-40,152){$\de m^2$}
\put(-40,127){$\De m^2$}
\put(-40,86){$\de \cri$}
\put(-60,30){$\De \cri$}
\end{picture}}
\caption{\label{expts2}
Lines of limiting sensitivity for
$\de m^2 \sim 10^{-5}$ eV$^2$,
$\De m^2 \sim 10^{-3}$ eV$^2$,
$\de\cri \sim 10^{-22}$,
$\De\cri \sim 10^{-17}$.
The shaded regions are those of Fig.\ 1.}
\end{figure}

A particularly simple FC model consists of a single
nonzero coefficient matrix such as $(c_L)^{TT}_{ab}$.
Some features of this model
are similar to the conventional massive-neutrino case,
but there is unusual energy dependence.
Here,
we take advantage of this energy dependence
to illustrate one type of mechanism through which
Lorentz violation might provide a solution to the LSND anomaly.

Lines of limiting sensitivity for the two mass-squared differences 
$\de m^2$ and $\De m^2$ 
used in the conventional global analysis 
are shown in Fig.\ \ref{expts2}.
The mixing angles are such that
$\nu_e$ oscillations are almost completely controlled by $\de m^2$.
Therefore, 
one can expect to see only $\nub_e$ mixing in KamLAND,
in solar neutrino experiments,
and possibly in the lowest-energy atmospheric-neutrino experiments.
The observed atmospheric oscillations are due to $\De m^2$,
which controls $\nu_\mu\mix\nu_\ta$ mixing.
Since LSND lies well below both the $\de m^2$ and the $\De m^2$ lines,
no oscillations are predicted.

Replacing the mass-squared differences $\de m^2$ and $\De m^2$
with a nonzero coefficient matrix $(c_L)^{TT}_{ab}$ 
produces an effective hamiltonian $\heff$ that can be parametrized 
as described in Appendix \ref{smeterms},
using two eigenvalue differences 
and CKM-like mixing angles and phases.
For simplicity, 
we choose here to mimic the usual solution 
by taking vanishing phases and $\th^{TT}_{13}$,
and we consider only the case $\th^{TT}_{23}=\pi/4$.
This leaves three degrees of freedom:
two eigenvalue differences,
and one mixing angle $\th^{TT}_{12}\equiv\th$.
It turns out to be convenient to work with 
two linear combinations of the eigenvalue differences,
defined by 
\bea
\de \cri&\equiv&\frac43\big((c_L)^{TT}_{(3)}-(c_L)^{TT}_{(2)}\big) ,
\nonumber \\
\De \cri&\equiv&\frac43\big((c_L)^{TT}_{(2)}-(c_L)^{TT}_{(1)}\big) .
\eea
The probabilities for this case are then
\bea
P_{\nu_e\to\nu_e}&=&
1-\sin^22\th\sin^2(\De\cri LE/2), \nonumber\\
P_{\nu_\mu\to\nu_\mu}&=&P_{\nu_\ta\to\nu_\ta}
=1-\frac14\sin^22\th\sin^2(\De\cri LE/2) \nonumber\\
&&-\sin^2\th\sin^2((\De\cri +\de\cri)LE/2) \nonumber\\
&&-\cos^2\th\sin^2(\de\cri LE/2),  \nonumber\\
P_{\nu_e\mix\nu_\mu}&=&P_{\nu_e\mix\nu_\ta}
=\half\sin^22\th\sin^2(\De\cri LE/2),  \nonumber\\
P_{\nu_\mu\mix\nu_\ta}
&=&-\frac14\sin^22\th\sin^2(\De\cri LE/2) \nonumber\\
&&+\sin^2\th\sin^2((\De\cri +\de\cri)LE/2) \nonumber\\
&&+\cos^2\th\sin^2(\de\cri LE/2) .
\label{probsm2}
\eea
The corresponding antineutrino expressions
are identical.

A possible approach is illustrated in the figure.
The line of sensitivity
for the larger difference $\De\cri$ 
can be chosen to lie just above CHOOZ and LSND.
This produces only a small effect in these experiments 
and may provide an explanation
for LSND that may not conflict with CHOOZ.
The remaining difference $\de\cri$ can
then be chosen to explain atmospheric data.
The above situation somewhat resembles the conventional mass solution,
with the role of $\de m^2/2E$ replaced by $\De\cri E$ 
and that of $\De m^2/2E$ replaced by $\de\cri E$.
The angle $\th$ is the analogue of
the solar-neutrino mixing angle.
However,
the energy dependences of the two cases differ substantially,
as is also evident from the figure.

To explore quantitatively how this approach might work,
consider the positive LSND and KamLAND results.
KamLAND detects $\nub_e$ from distant reactors
and found about a $61\%$ reduction in the flux.
Most reactors are 138-214 km from the detector,
and the corresponding $\nub_e$ energies 
fall in the range 1 MeV $\lsim E \lsim$ 10 MeV.
If KamLAND lies well above the $\De\cri$ line,
the relevant quantity is the average survival probability
$\vev{P_{\nub_e\to\nub_e}} =1-\half\sin^22\th\simeq 61\%$,
yielding a mixing angle given by $\sin^22\th \simeq 0.78$.
Also,
assuming LSND is in a region of small oscillation effects,
then we can approximate
$P_{\nub_\mu\to\nub_e}\approx
\half\sin^22\th(\De\cri LE/2)^2 \simeq 0.26\%$.
Then,
for $E\simeq 45$ MeV and $L\simeq 30$ m
we obtain $\De\cri \simeq 2.4\times 10^{-17}$.
Thus, 
in this simple scenario,
these two experiments suggest coefficient values 
near $\sin^22\th \simeq 0.78$ and
$\De\cri \simeq 2.4\times 10^{-17}$,
in agreement with the estimates of Sec.\ \ref{lsnd}.

The remaining coefficient $\de\cri$ can then be chosen 
to match observed atmospheric-neutrino effects.
The coefficient $\De\cri$ is relatively large in this region
and generates rapid oscillations.
Averaging over these for any value of $\de\cri$
leaves a muon-neutrino survival probability of
either
$P_{\nu_\mu\to\nu_\mu}\simeq 0.54-0.27\sin^2(\de\cri LE/2)$
or
$P_{\nu_\mu\to\nu_\mu}\simeq 0.77-0.73\sin^2(\de\cri LE/2)$,
depending on the solution for $\th$. 
Note that the latter expression resembles 
the usual maximal-mixing solution within an overall scale factor,
except for the unusual energy dependence in the oscillation length.

Interestingly,
atmospheric electron-neutrino oscillations
are present in this model but are largely unobserved
due to a compensation mechanism.
The averaged $\nu_e$ survival probability is 
$P_{\nu_e\to\nu_e}=61\%$,
as above,
and the $\nu_e\mix\nu_\mu$ mixing probability
is $P_{\nu_e\mix\nu_\mu}=19.5\%$.
The observed flux of atmospheric electron neutrinos
is a combination of the survival flux
and the appearance flux from mixing with muon neutrinos.
Since the ratio of muon neutrinos to electron neutrinos 
is approximately 2, 
the predicted effective flux of atmospheric electron neutrinos
is approximately $61\% + 2(19.5\%) \simeq 100\%$ of the flux
in the absence of oscillations,
in agreement with indications from existing data.
Essentially,
this compensation mechanism works
because the disappearance probability 
$1 - P_{\nu_e\to\nu_e}$
of electron neutrinos given by Eq.\ \rf{probsm2}
is a factor of two greater than 
the appearance probability $P_{\nu_e\mix\nu_\mu}$
of muon neutrinos from mixing,
resulting in no net suppression 
in the total observed electron-neutrino flux.

The compensation mechanism 
\it per se \rm
is independent of Lorentz violation
and can be applied whenever 
$1 - P_{\nu_e\to\nu_e}\approx 2P_{\nu_e\mix\nu_\mu}$,
including in the conventional massive case.
Note,
however,
that Monte Carlo calculations suggest the flux ratio 
increases dramatically above 2 for energies 
over about 10 GeV \cite{mc},
so the compensation mechanism is likely to fail at higher energies.
Note also that,
in the case of the above Lorentz-violating model,
the rapid oscillations at high energies 
also help to mask $\nu_e$ oscillations.
Although these rapid oscillations can change the overall flux, 
they also tend to smooth away the observable $E$ and $L$ dependences
that form the basis for some analyses. 

This simple model serves to illustrate a possible strategy
that might remedy the conflict
between LSND and reactor experiments,
but it may well introduce other conflicts 
between LSND and accelerator experiments testing 
$\nu_e\to\nu_\ta$ and $\nu_\mu\to\nu_\ta$
\cite{chorus,nomad1}
or $\nu_\mu\to\nu_e$
\cite{nutev,nomad2}.
Note also that some work has been done 
to check for unconventional energy dependences
in the atmospheric data 
\cite{efit},
suggesting that the usual energy dependence is preferred.
However, 
these analyses are limited to two generations 
and do not consider possible direction dependences 
or $\nu\mix\nub$ mixing.
A complete treatment would also need to include
the effects of the Earth's matter potential,
which introduces additional energy dependence.
The point is that $G_Fn_e \sim 10^{-22}$ GeV for the Earth, 
and at atmospheric-neutrino energies
this is comparable to the contribution from $\de\cri$
shown in Fig.\ \ref{expts2}.
In any case,
interesting sensitivities to Lorentz violation
could be achieved with a complete analysis of existing data.

\subsubsection{Example:
  $(a_L)^T_{e\mu}\ne 0$, $(c_L)^{TT}_{\mu\ta}\ne 0$}
\label{acmod}

We turn next to an FC model with mixed energy dependence,
incorporating only two nonzero coefficients 
$(a_L)^T_{e\mu}\equiv \aem$
and
$\frac43(c_L)^{TT}_{\mu\ta}\equiv \cmt$
and no mass terms.
This case includes both Lorentz and CPT violation
but remains rotation symmetric.
The presence of both a dimensionless coefficient
and a dimension-one coefficient
leads to unusual energy behavior 
in the vacuum-mixing angles as well as the oscillation lengths.
This contrasts with the previous case,
in which only the oscillation lengths have unconventional 
energy dependence.
Note that both $\aem$ and $\cmt$ are arbitrary 
to an unobservable phase,
and therefore they can be taken real and nonnegative 
without loss of generality.

The behavior in this model can be understood qualitatively as follows.
At low energies $E\ll \aem / \cmt$ 
relative to the critical energy $\aem / \cmt$,
the $\aem$ term dominates the effective hamiltonian.
As a result, 
$\nu_\ta$ decouples from $\nu_e$ and $\nu_\mu$,
so only $\nu_e\mix\nu_\mu$ mixing occurs.
In contrast,
for high energies $E\gg \aem / \cmt$,
$\cmt$ dominates and only $\nu_\mu\mix\nu_\ta$ mixing occurs.
At intermediate energies $E\sim \aem / \cmt$,
the two terms are comparable
and produce complicated energy dependence 
with mixing between all three neutrinos.

This behavior is similar to the observed energy dependence
in the solar-neutrino flux.
In the usual analysis with massive neutrinos,
the observed energy dependence is explained through matter effects.
However,
the same type of behavior can appear in Lorentz-violating scenarios 
even without matter.
To demonstrate this,
we need the probabilities for the current model:
\bse\bea
P_{\nu_e\to\nu_e}&=&
1-4\sin^2\th \cos^2\th\sin^2(\pi L/L_0) \nonumber \\
&&-\sin^4\th\sin^2(2\pi L/L_0) , \label{pee} \\
P_{\nu_\mu\to\nu_\mu}&=&
1-\sin^2(2\pi L/L_0) , \label{pmm} \\
P_{\nu_\ta\to\nu_\ta}&=&
1-4\sin^2\th \cos^2\th\sin^2(\pi L/L_0) \nonumber \\
&&-\cos^4\th\sin^2(2\pi L/L_0) , \label{ptt} \\
P_{\nu_e\mix\nu_\mu}&=&
\sin^2\th\sin^2(2\pi L/L_0) , \label{pem} \\
P_{\nu_e\mix\nu_\ta}&=&
\sin^2\th \cos^2\th\big(4\sin^2(\pi L/L_0)\nonumber\\
&&-\sin^2(2\pi L/L_0)\big) , \label{pet} \\
P_{\nu_\mu\mix\nu_\ta}&=&
\cos^2\th\sin^2(2\pi L/L_0) , \label{pmt}
\eea\label{p}\ese
where
\bea
\sin^2\th&=&\aem^2/(\aem^2+\cmt^2E^2) ,
\nonumber\\
2\pi/L_0&=&\sqrt{\aem^2+\cmt^2E^2} .
\eea
The antineutrino probabilities are
again identical since the quantities
$\sin^2\th$ and $L_0$ are symmetric
under $\aem \to -\aem$.
We remark in passing that this model serves as an example 
in which CPT is violated 
but the traditional test of CPT discussed in Sec.\ \ref{cptprop} 
fails as an indicator of the CPT violation.

\begin{figure}
$\amt E/\mem^2$
\centerline{
\psfig{figure=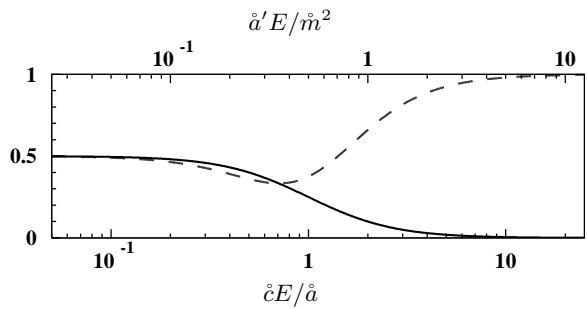,width=0.9\hsize}}
$\cmt E/\aem$
\caption{\label{prob}
Solar-neutrino survival
probability assuming adiabatic
propagation (solid),
and average survival probability
for vacuum oscillations (dashed).}
\end{figure}

The solar-neutrino vacuum-oscillation survival probability 
is given by Eq.\ \rf{pee}.
As usual,
depending on the size of the coefficients,
matter effects can drastically alter the survival rates.
Consider, 
for example, 
a simple matter-dominated case where the matter potential 
at the point of $\nu_e$ production dominates $\heff$.
Assuming adiabatic propagation,
neutrinos are produced 
in the highest-eigenvalue state of $\heff(R\simeq 0)$
and emerge from the Sun 
in the highest-eigenvalue state of $\heff(R= R_\odot)$.
The overlap between this state and an electron-neutrino state
is proportional to $\sin\th/\sqrt{2}$.
Consequently,
the average survival probability
for the matter-dominated case in an adiabatic approximation is
\beq
\vev{P_{\nu_e\to\nu_e}}_{\rm adiabatic}
=\half\sin^2\th .
\label{adiabatic}
\eeq
In contrast, 
the average for the case where matter effects can be neglected is 
\beq
\vev{P_{\nu_e\to\nu_e}}_{\rm vacuum}
=1-2\sin^2\th+\frac32\sin^4\th.
\label{vacuum}
\eeq
These probabilities are plotted on Fig.\ \ref{prob} 
as a function of energy in units of $\aem/\cmt$.

The observed flux is consistent with the figure,
since low-energy experiments suggest 
an approximate survival probability of 1/2
\cite{gno,gallex,sage},
while higher-energy experiments favor about 1/3
\cite{homestake,sno,sksol}.
Note that both cases shown in Fig.\ \ref{prob} 
yield an average survival probability of 1/3
at $E=\aem/\sqrt{2}\cmt$.
By choosing the ratio $\aem/\cmt$ to coincide
with the peak of the solar $^8$B spectrum
($E_{\rm peak}\simeq 6.4$ MeV),
this simple massless Lorentz- and CPT-violating model 
can be made to reproduce the gross features 
of the observed solar-neutrino flux.
This corresponds to imposing $\aem/\cmt\simeq 9$ MeV.

The above discussion only depends on the ratio of coefficients.
To get a sense of the size of coefficients required in a realistic case,
we can consider what KamLAND implies for $\aem$ and $\cmt$.
Taking a representative neutrino 
to have energy $E=5$ MeV and baseline $L=200$ km
and assuming that it oscillates no more than once,
the ratio $\aem/\cmt\simeq 9$ MeV
and the survival probability $P_{\nub_e\to\nub_e}\simeq 61\%$ 
can be used to extract approximate values
$\aem\simeq 7\times 10^{-22}$ GeV
and
$\cmt\simeq 8\times 10^{-20}$.
The lines of sensitivity for these values on Fig.\ \ref{expts}
are approximately $L\sim 10^{21}$ GeV$^{-1}$ and $LE\sim 10^{19}$,
passing just above KamLAND and intersecting in the solar-energy region,
thereby producing the energy dependence seen in Fig.\ \ref{prob}.

\subsubsection{Example: 
  $(\mt)_{e\mu}\ne 0$, $(a_L)^{TT}_{\mu\ta}\ne 0$}\label{mamod}

As a variation on the above model,
we next consider a special FC case with nonzero mass
$(\mt)_{e\mu}\equiv\mem^2$ and coefficient
$(a_L)^{TT}_{\mu\ta}\equiv \amt$ for Lorentz and CPT violation.
This model has many qualitative features of the previous one.
At small energies,
the mass $\mem$ controls mixing between $\nu_e$ and $\nu_\mu$,
while at large energies 
$\amt$ dominates and produces
mixing between $\nu_\mu$ and $\nu_\ta$.

The probabilities for this model are given by Eqs.\ \rf{p},
\rf{adiabatic}, and \rf{vacuum},
but with the definitions
\bea
\sin^2\th&=&\mem^4/(\mem^4+4\amt^2 E^2) ,
\nonumber \\
2\pi/L_0&=&\sqrt{(\mem^2/2E)^2+\amt^2} .
\eea
The analysis of this model parallels the previous case.
Indeed,
Fig.\ \ref{prob} also holds
for the solar-neutrino probabilities 
in terms of $\mem$ and $\amt$,
using the scale shown on the top axis.
Applying the same arguments as before yields
the ratio $\mem^2/\amt\simeq 18$ MeV
and candidate values
$\mem^2\simeq 7\times 10^{-6}$ eV$^2$ and
$\amt\simeq 4\times 10^{-22}$ GeV.

A key difference between this case
and the previous $\aem$-$\cmt$ model 
is the asymptotic behavior of the oscillation length.
In the $\aem$-$\cmt$ case,
$L_0\to 2\pi/(\cmt E)$ at high energies.
In contrast, 
the oscillation length in the present $\mem$-$\amt$ model
approaches a constant at high energies,
$L_0\to 2\pi/\amt$.
Consider the consequences for atmospheric neutrinos.
Note that in the high-energy limit of both cases, 
$\sin^2\th\to0$ and so $P_{\nu_e\to\nu_e}\to0$, 
in agreement with observation.
However,
the first model with $\cmt\simeq 8\times 10^{-20}$
gives $L_0\simeq 2\pi/(\cmt E) \simeq (15$ km GeV$)/E$,
whereas the second model with $\amt\simeq 4\times 10^{-22}$ GeV
yields $L_0\simeq 3100$ km.
These differ from the usual massive-neutrino explanation 
of the atmospheric data,
which has $\De m^2\simeq 3\times 10^{-3}$ eV$^2$
and results in $L_0=4\pi E/\De m^2\simeq 800E$ km/GeV.

We emphasize that both this special model and the previous one 
involve only two degrees of freedom,
whereas the usual massive-neutrino solution requires 
two mass-squared differences and at least two mixing angles.
Including additional coefficients for Lorentz violation  
can only add flexibility to the analysis.
For example, 
one might consider a combination of the two examples above,
which would have four degrees of freedom.
With additional freedom,
it seems likely that an appropriate simple Lorentz-violating scenario
could be constructed that would reproduce most oscillation data.
This also suggests that existing data analyses
appear insufficient to exclude many forms of Lorentz and CPT violation,
or even to distinguish between oscillations due to mass 
and those due to Lorentz violation.

\subsection{Direction-dependent and
$\nu\mix\nub$ mixing models}\label{dirmods}

Lorentz violation naturally allows directional dependence 
in oscillation parameters through the violation of
rotation invariance.
An interesting subset of direction-dependent models are
those involving $\nu\mix\nub$ mixing 
via nonzero $g^{\mn\si}$ and $H^\mn$ coefficients
in the theory \rf{heff}.
In the general case,
nonzero $\nu\mix\nub$ mixing represents one way
to generate as many as five distinct oscillation lengths 
without incorporating sterile neutrinos.
However,
we limit attention in this subsection 
to a simple model that reveals some key features
of $\nu\mix\nub$ mixing.
For illustrative purposes, 
it suffices to consider mixing in only one neutrino species,
say $\nu_e\mix\nub_e$.
This case may nonetheless have physical relevance,
since it implies significant effects on reactor experiments 
and solar neutrinos
and might possibly also shed light on the LSND anomaly.

\subsubsection{General one-species model}\label{1spmod}

The restriction to the two-dimensional $\nu_e$-$\nub_e$ subspace 
radically simplifies the form of the effective hamiltonian \rf{heff}.
Since the coefficients $(\mt)_{ee}$ and $(c_L)_{ee}$ are real,
they lead to terms proportional to the identity
that have no effect on oscillatory behavior
and can therefore be ignored.
Moreover,
Eq.\ \rf{C} implies that $(H^\mn\cmat)_{ab}$ 
is antisymmetric in generation space, 
so $(H^\mn\cmat)_{ee}=H^\mn_{ee^C}=0$.
Therefore,
the most general single-flavor theory without mass differences
is given by a $2\times 2$ effective hamiltonian
containing only the coefficients
$(a_L)_{ee}^\mu$ and $(g^{\mn\si}\cmat)_{ee}=g^{\mn\si}_{ee^C}$
for Lorentz violation.
Note that both these terms are CPT odd.

For this general single-flavor model,
the probabilities are identical in form 
to those of the usual two-generation mixing case:
\bea
P_{\nu_e\mix\nub_e}&=&
1-P_{\nu_e\to\nu_e}=
1-P_{\nub_e\to\nub_e}
\nonumber \\
&=&\sin^22\th\sin^2 2\pi L/L_0 .
\eea
However, 
the mixing angle and oscillation length 
can have nontrivial 4-momentum dependence.
They are given by the expressions
\bea
\left(\fr{2\pi}{L_0}\right)^2&=&
\fr{|(a_L)_{ee}^\mu p_\mu|^2}{|\vp|^2}
+ |\sqrt{2}(\ep_+)_\nu p_\si \gt_{ee^C}^{\nu\si}|^2 ,
\nonumber \\
\sin^22\th&=&
\left(1+\fr{|(a_L)_{ee}^\mu p_\mu|^2}
{|\vp|^2|\sqrt{2}(\ep_+)_\nu p_\si \gt_{ee^C}^{\nu\si}|^2}
\right)^{-1} .
\eea
Note that these can also be written directly 
in terms of the neutrino-propagation angles $\Th$ and $\Ph$ 
defined in Eq.\ \rf{vectors}.

\subsubsection{Example:
  $\gt_{ee^C}^{ZT}\ne 0$}\label{gmod}

As an explicit example,
we consider a maximal-mixing special case
of the general single-flavor model
for which the only nonzero coefficient 
is $\gt_{ee^C}^{ZT}\equiv\Gc$.
In terms of the propagation angles $\Th$ and $\Ph$,
the oscillation length is found to be 
\beq
2\pi/L_0=|E\sin\Th\Gc|,
\eeq
and the mixing angle is $\sin^22\th=1$.
As in the previous examples, 
this case has unconventional energy dependence,
but unlike previous examples
it includes neutrino-antineutrino mixing
and also dependence on the direction of propagation 
through the propagation angle $\Th$.

To illustrate the effects of the direction dependence,
consider atmospheric neutrinos detected in the SK detector.
Neutrinos that enter the detector 
from the celestial north or south have $\sin\Th=0$ 
and therefore do not oscillate.
In contrast, 
neutrinos propagating in the plane parallel 
to the Earth's equatorial plane 
have $\sin\Th=1$ and experience maximal mixing
\cite{fn2}.
Analyses of SK data often neglect 
the difference between $\nu_e$ and $\nub_e$,
so they may be insensitive to this effect
because the total flux of electron neutrinos and antineutrinos 
is unchanged.
However, 
the same type of directional dependence 
can arise in more complicated scenarios 
with $\nu_e\mix\nu_\mu\mix\nu_\ta$ mixing,
and this could drastically affect 
the up-down asymmetry measurements of SK.

As another example consider KamLAND,
which detects neutrinos from several reactors at different locations.
The total flux $\ph_{\rm total}(E)$ of $\nub_e$ can be written 
\beq
\ph_{\rm total}(E)
=\sum_j \ph_j(E) P_{\nub_e\to\nub_e}(E,L_j,\Th_j),
\eeq
where the $\ph_j(E)$ are the fluxes from the
individual reactors in the absence of oscillations,
and $\Th_j$ are appropriate propagation angles
determined by the relative positions of the reactors 
and the KamLAND detector.
We can approximate the positions of the reactors 
as being located in the plane tangent to the surface of the Earth 
at the location of the detector.
It follows that neutrinos from reactors 
positioned directly north and south of the detector have
$\Th_j\simeq180^\circ-\ch$ and $\Th_j\simeq\ch$,
where $\ch\simeq 36^\circ$
is the latitude of the detector.
In contrast, 
neutrinos arriving from the east or west have $\Th_j\simeq 90^\circ$.
This results in an approximate allowed range for the $\Th_j$
given by $\sin^2\Th_j\gsim \sin^2\ch$,
implying that the $\nub_e$ from every reactor 
experience some degree of oscillation 
on their way to the KamLAND detector.
However,
the net result differs from the flux
in a comparable rotation-symmetric model 
with a dimensionless coefficient.

For solar neutrinos,
the allowed range for $\Th$ is given by
$\sin^2\Th\gsim\cos^2\et\simeq0.85$
because the Earth's orbital and equatorial planes differ by
approximately $\et=23^\circ$.
The true value of $\sin^2\Th$ oscillates between $\sin^2\Th=1$
in the spring or fall
and $\sin^2\Th\simeq0.85$ in the summer or winter.
This simple model therefore predicts a semiannual variation 
in the solar-neutrino data.

As suggested in Sec.\ \ref{lsnd},
oscillations of $\nu_e$ into $\nub_e$ 
may provide an alternative approach 
to resolving the LSND anomaly.
If the LSND result is reinterpreted
as an oscillation of $\nu_e$ into $\nub_e$,
then the transition probability is 
likely to be comparable to the reported value of about $0.26\%$
because roughly equal numbers of $\nu_e$ and $\nub_\mu$ are produced.
Since mixing in this model is caused 
by the dimensionless coefficient $\Gc$,
a reasonable strategy here is similar to  
that adopted for the $\de\cri$-$\De\cri$ model
in Sec.\ \ref{dcmod},
where a dimensionless coefficient is chosen 
to have its line of sensitivity just above CHOOZ and LSND
in Fig.\ \ref{expts}.
This causes a small oscillation in LSND
but avoids the null constraints from reactor experiments.
Taking the energy of a typical $\nu_e$ to be about $E=35$ MeV
and the distance to be $L=30$ m in LSND,
and assuming that the small transition probability 
is due to a small $L/L_0$,
we can write 
$P_{\nu_e\to\nub_e} =\sin^22\th\sin^22\pi L/L_0
\simeq(\sin\Th\Gc LE)^2 \simeq 0.26\%$.
For LSND, 
the detector is situated approximately to the east of the source.
This implies that the angle between celestial north 
and the direction of propagation of the neutrinos is near $90^\circ$,
which results in the estimate $|\Gc| \simeq 10^{-17}$.

In contrast, 
the KARMEN detector is located roughly to the south 
of the neutrino source,
at latitude $\ch \simeq 51^\circ$.
We can therefore approximate
$\Th \simeq 180^\circ - \ch \simeq 129^\circ$.
Taking $E=35$ MeV and $L=18$ m for KARMEN 
yields a transition probability
$P_{\nu_e\to\nub_e} =\sin^22\th\sin^22\pi L/L_0
\simeq(\sin\Th\Gc LE)^2 \simeq 0.06\%$.
This is more than four times smaller than the LSND probability
as a consequence of the different propagation direction 
and the smaller distance,
confirming that direction dependence 
could help reconcile the apparent conflict between KARMEN and LSND.

In the above model, 
the directional dependence is rather limited
because the coefficient $\Gc$ introduces only $\Th$ dependence.
This causes minimal variation for any experiments
with both neutrino source and detector fixed on the Earth's surface,
since the angle $\Th$ is fixed as the Earth rotates
and is therefore a constant experiment-dependent quantity.
However,
other coefficients can produce a strong dependence on $\Ph$ as well.
For instance, 
suppose we choose $\gt_{ee^C}^{ZX}$ 
instead of $\gt_{ee^C}^{ZT}$.
The result is an oscillation length given by
$2\pi/L_0=|E\sin^2\Th\cos\Ph \gt_{ee^C}^{ZX}|$.
The dependence on $\Ph$ can substantially 
change the nature of an experiment.
For purely terrestrial experiments,
where the source and detector are fixed to the surface of the Earth,
it follows that $\Ph = \om_\oplus (T-T_0)$,
where $\om_\oplus\simeq 2\pi/$(23 h 56 min)
is the Earth's sidereal frequency 
and $T_0$ is an appropriately chosen experiment-dependent offset.
For solar neutrinos, 
$\Ph$ varies as the Earth orbits the Sun,
$\Ph\approx \Om_\oplus (T-T_0)$,
where $\Om_\oplus = 2\pi/$(1 year).

\subsection{Lorentz-violating seesaw models}\label{ssmods}

The above models demonstrate 
some of the striking behavior at different energy scales
that can arise from Lorentz and CPT violation.
Mixed energy dependence 
among the coefficients for Lorentz violation in $\heff$ 
can also lead to a Lorentz-violating seesaw mechanism
that occurs without mass and only in particular energy regimes.
This can lead to counterintuitive phenomena,
such as the appearance of a pseudomass 
in the bicycle model of Ref.\ \cite{nu}.
In this model,
an oscillation length emerges at high energies 
that behaves like a mass-squared difference,
even though no mass-squared differences exist in the theory.

\begin{figure}
\centerline{\psfig{figure=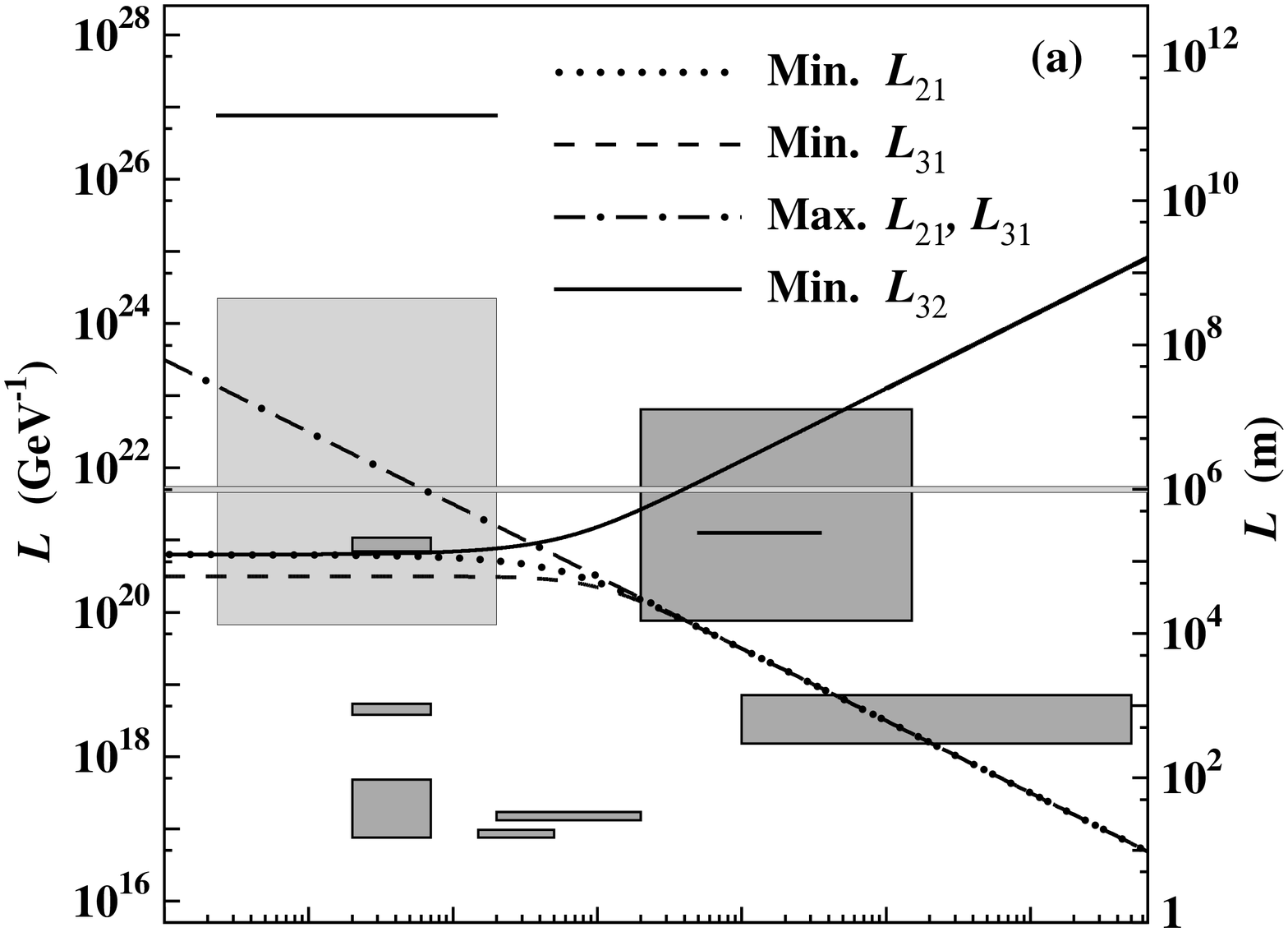,width=\hsize}}
\centerline{\psfig{figure=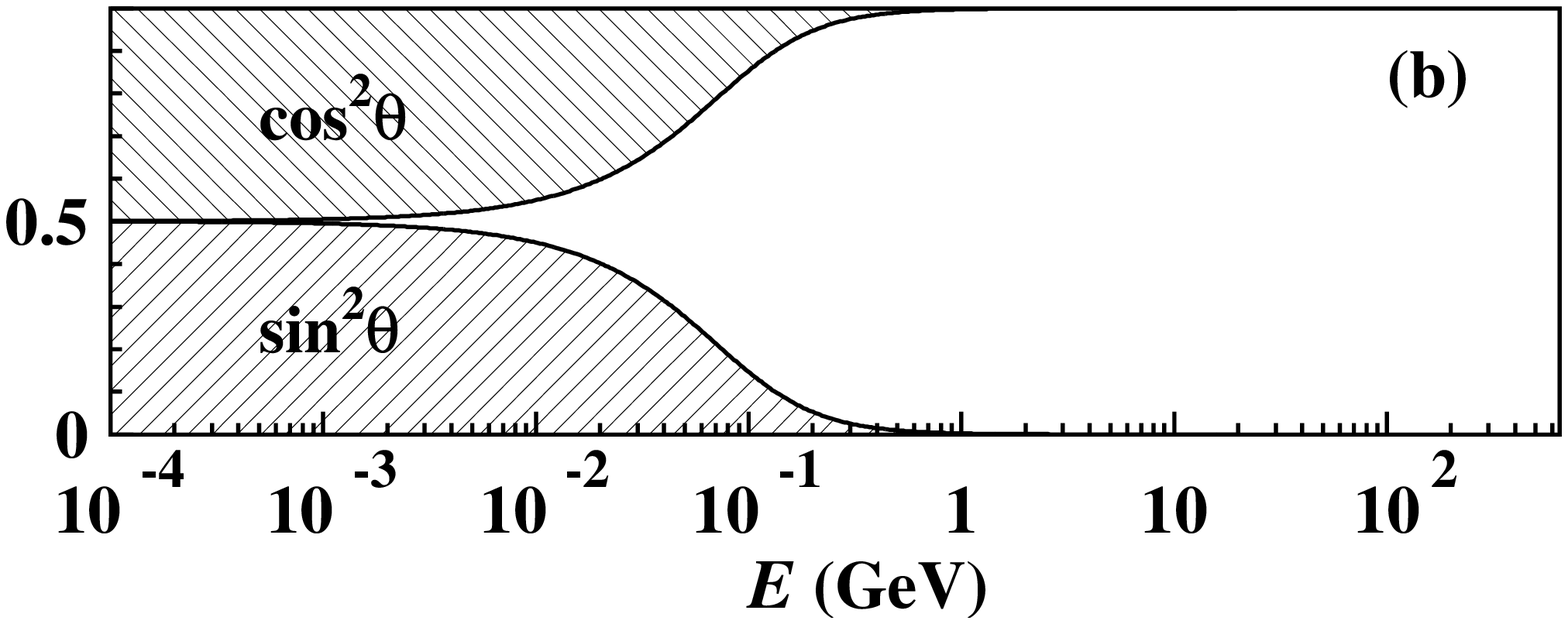,width=\hsize}}
\caption{\label{caexpts}
Range of oscillation parameters versus energy 
in the bicycle model with 
$\cee=10^{-19}$ and $\a3em=10^{-20}$ GeV.
(a) Minimum ($\cos^2\Th=1$)
and maximum ($\cos^2\Th=0$)
of the various oscillation lengths
$L_{ab}\equiv2\pi/\De_{ab}$.
Note that $L_{32}$ is unbounded.
(b) The allowed range of $\sin^2\th$ and $\cos^2\th$
over all possible directions,
$0\le\cos^2\Th\le1$,
as a function of energy.}
\end{figure}

The bicycle model has nonzero coefficients
$\fr43(c_L)^{TT}_{ee} =\fr43(c_L)^{JJ}_{ee} \equiv 2\cee$
and
$(a_L)^Z_{e\mu}=(a_L)^Z_{e\ta}\equiv\a3em/\sqrt{2}$.
The basic behavior of the oscillation lengths 
$L_{ab}\equiv2\pi/\De_{ab}$
and the energy-dependent mixing angle $\th$
are illustrated in Fig.\ \ref{caexpts}.
A key feature is that at high energies
the line associated with the oscillation length $L_{32}$ 
resembles that from a nonzero mass-squared difference.
It turns out that the resulting high-energy dynamics
reduces to two-generation maximal mixing,
$P_{\nu_\mu\mix\nu_\ta}\simeq\sin^2(\De m^2_\Th L/4E)$,
with a Lorentz- and CPT-violating pseudomass 
$\De m^2_\Th=\a3em^2\cos^2\Th/\cee$.

Unexpected effects of this type can be expected 
whenever the low- or high-energy limit of $\heff$ 
contains degeneracies.
Consider, for example,
a $3\times 3$ hamiltonian $\heff$ for which there exists a basis,
not necessarily the flavor basis,
in which we can write 
\beq
\heff = \left(\begin{array}{ccc}
2h_1   & h_2 & h_3 \\
h^*_2 & 0 & 0   \\
h^*_3 & 0   & 0
\end{array}\right) ,
\eeq
where irrelevant diagonal terms are neglected.
The interesting eigenvalue difference for this case is
$\De = \sqrt{(h_1)^2+|h_2|^2+|h_3|^2}-h_1$.
Suppose that the mixed energy dependence introduced 
by combinations of masses,
dimension-one coefficients,
and dimensionless coefficients enforces 
$h_1\gg\sqrt{|h_2|^2+|h_3|^2}$ 
at some energy scale.
Expanding the eigenvalue difference then yields
\cite{fn3}
$\De \approx \half(|h_2|^2+|h_3|^2)/h_1+\cdots$.

In the bicycle model,
$h_2$ and $h_3$ arise from a dimension-one coefficient 
and are therefore constant with energy,
but $h_1$ arises from a dimensionless coefficient 
and therefore grows linearly with energy.
As a result, 
at high energies the eigenvalue difference 
is proportional to $E^{-1}$,
which resembles the usual mass case.
Using different combinations of masses 
and coefficients for Lorentz violation,
it is straightforward to construct similar models 
that produce $E^{-1}$, $E^{-2}$, or $E^{-3}$ dependence 
at high energies,
or $E^1$, $E^2$, or $E^3$ dependence at low energies.
More complicated $E^n$ dependences are possible 
when the full $6\times 6$ effective hamiltonian \rf{heff}
with $\nu\mix\nub$ mixing is considered.


\section{Discussion}\label{disc}

In this paper,
we have presented a general framework for the study
of Lorentz violation in the neutrino sector.
The key result is Eq.\ \rf{heff},
which represents the general effective hamiltonian $\heff$ 
for neutrino propagation in the presence
of Lorentz and CPT violation.
We have extracted theoretical implications of this hamiltonian
and have initiated a study of experimental sensitivities
to the predicted effects.
The various simple models of Sec.\ \ref{models}
illustrate some of the key physical features
and offer numerous options for future exploration.

Our analysis shows that 
the data from existing and near-future neutrino experiments
could be used to attain interesting sensitivities
to possible Lorentz-violating effects.
Moreover,
the existing analyses appear insufficient to
exclude the possibility that some 
or perhaps even all the established neutrino-oscillation signals 
are due to Lorentz violation.

An interesting open theoretical challenge is to identify
from the plethora of available choices
one or more elegant models with features compatible with observed data,
preferably involving only a small number of degrees of freedom.
One simple candidate is the bicycle model
\cite{nu},
which has no mass-squared differences and only two degrees of freedom
rather than the four used in the conventional massive-neutrino analysis, 
but which nonetheless reproduces the major observed features of
neutrino behavior. 
This and similar models offer one possible path to explore,
but it is likely that many other qualitatively different
and interesting cases exist. 

On the experimental front,
confirming or disproving these ideas
would involve analysis of existing and future data
to seek a `smoking-gun' signal for Lorentz violation.
In the remainder of this section,
we summarize some possible smoking-gun signals
and then offer some remarks about experimental prospects
for detection of Lorentz and CPT violation.

\subsection{Generic predictions}\label{predictions}

The numerous options for coefficients for Lorentz and CPT violation
and the size of unexplored $L$ versus $E$ space
are impediments to a completely general analysis.
An alternative strategy to uncover evidence of Lorentz violation
is to seek model-independent features 
that represent characteristic signals.
We list here six classes of signal.
Confirmed observation of any of them  
would be evidence supporting the existence of Lorentz violation.

\begin{trivlist}
\item
{\it Class I:\ Spectral anomalies.}
Each coefficient for Lorentz violation
introduces energy dependence differing from the usual case.
Detection of a vacuum oscillation length
that is constant in $E$
or inversely proportional to $E$ to some power
would constitute a clear signal of Lorentz violation.
Note that combinations of 
masses, dimension-one coefficients,
dimensionless coefficients, and matter potentials
can produce more complicated energy dependences
in both oscillation lengths and mixing angles.
In general,
a mixing angle is constant in energy
only if all relevant coefficients for Lorentz violation,
masses, and matter effects have the same dimension,
which requires no more than one of these to be present.

\item
{\it Class II:\ $L$--$E$ conflicts.}
This class of signal refers to any null or positive measurement
in a region of $L$--$E$ space that conflicts with
all scenarios based on mass-squared differences.
For example,
consider a solid line in Fig.\ \ref{expts} 
passing through CHOOZ.
A measurement of substantial oscillation in the $\nub_e$ sector 
in any experiment below this line would be in direct conflict 
with a mass-based interpretation of the CHOOZ results.
Signals in this class might best be sought 
by searching for oscillation effects
in each species of neutrino and antineutrino
for regions of $L$--$E$ space 
in which conventional oscillations are excluded.
Of the six classes of signal discussed in this section, 
this is the only one 
for which there is presently some positive evidence,
the LSND anomaly.

\item
{\it Class III:\ Periodic variations.}
This class involves signals for rotation-invariance violations
and contains two subclasses:
sidereal variations and annual variations.
Consider first sidereal variations,
which have been widely adopted as the basis 
for Lorentz-violation searches in other sectors of the SME.
In terrestrial experiments
with both the detector and the source fixed on the Earth,
the direction of neutrino propagation relative to 
the Sun-centered frame changes during the sidereal day
due to the rotation of the Earth.
The induced periodic variation of observables with time 
represents a signature of Lorentz violation.
In the Sun-centered frame, 
the neutrino-propagation angle $\Th$ is constant for a fixed source,
but the angle $\Ph$ varies periodically 
according to $\Ph=\om_\oplus (T-T_0)$,
where $T_0$ is an experiment-dependent time 
at which the detector and source both lie in a plane 
parallel to the $XZ$ plane with the detector at larger values of $X$.
The resulting neutrino-oscillation probabilities 
exhibit periodic variations
at multiples of the sidereal frequency $\om_\oplus$.
The second class of periodic signals,
annual variations,
can also arise directly from rotation-invariance violation.
For solar-neutrino experiments,
the source is the Sun and the detector changes location with time  
as a consequence of the orbital motion of the Earth about the Sun. 
One can therefore expect variations at the Earth orbital frequency 
$\Om_\oplus$ and its harmonics.
In this context,
note that the direction $\hat p$ of solar neutrino propagation 
in the Sun-centered frame is uniquely given by
$\hat p=(-\cos\Om_\oplus T, 
-\cos\et\sin\Om_\oplus T,
-\sin\et\sin\Om_\oplus T)$,
where $\et\simeq 23.4^\circ$
is the angle between the
Earth's equatorial and orbital planes.
We remark in passing that suppressed annual variations can also arise
indirectly as boost-violating effects 
\cite{cavexpt,km,spaceexpt,cane}
in experiments with terrestrial 
and possibly atmospheric neutrino sources,
as a result of the noninertial nature of the Earth's motion
around the Sun.

\item
{\it Class IV:\ Compass asymmetries.}
This class also results from rotation-invariance violations,
but the signals are independent of time.
They can be characterized as the observation
of unexplained directional asymmetries
at the location of the detector.
For terrestrial and atmospheric experiments,
averaging over time eliminates the dependence 
on the neutrino-propagation angle $\Ph$,
so the result depends only on energy and the angle $\Th$.
Rotation-symmetry violations can therefore cause 
a difference in observed properties of neutrinos 
originating from different directions.
Note that the east and west directions are equivalent 
under the averaging process,
since the $\Ph$ dependence is eliminated,
but direct comparison of the north, south, and east directions
would be of interest for these signals.
Note also that the $\Th$ dependence 
typically introduces vertical up-down effects
and could include,
for example,
modifications in the up-down asymmetry of atmospheric neutrinos.
We remark also that compass asymmetries 
can carry information completely independent of 
the information in periodic variations.
This is seen in the example in Sec.\ \ref{gmod},
which has $\Th$ dependence but no $\Ph$ dependence
and consequently predicts compass asymmetries 
without sidereal variations.

\item
{\it Class V:\  Neutrino-antineutrino mixing.}
This class of signal includes any appearance measurement 
that can be traced to $\nu\mix\nub$ oscillation.
Any model with nonzero coefficients of type $g$ or $H$
exhibits this behavior,
including the class of simple one-species models
discussed in Sec.\ \ref{1spmod}.
Note that this class of signal involves lepton-number violation.

\item
{\it Class VI:\ Classic CPT test:
  $P_{\nu_b\to\nu_a}\ne P_{\nub_a\to\nub_b}$.}
This is the traditional test of CPT
discussed in Sec.\ \ref{cptprop},
involving violation of the result \rf{cpt1}.
A related signal would be violation of the second result,
Eq.\ \rf{cpt2},
which also involves $\nu\mix\nub$ mixing.

\end{trivlist}

\subsection{Experimental Prospects}\label{prospects}

\begin{table*}
\renewcommand{\arraystretch}{1.5}
\begin{tabular}{ccc||ccccc}
\hline
\hline
\multicolumn{3}{c||}{Coefficients}&
\multicolumn{5}{c}{Estimated sensitivities from Fig.\ \ref{expts}} \\
\hline
\makebox[110pt][c]{Matrix}&
\makebox[30pt][c]{DOF}&
\makebox[80pt][c]{Signal classes}&
\makebox[50pt][c]{Solar}&
\makebox[50pt][c]{Atmospheric}&
\makebox[50pt][c]{Reactor}&
\makebox[50pt][c]{Short base.}&
\makebox[50pt][c]{Long base.}\\
\hline
$(a_L)^T$&8&{\it I,II,VI}&
$-27$ & $-23$ & $-21$ & $-19$ & $-21$ \\
$(a_L)^J$&24&{\it I,II,III,IV,VI}&
$-27$ & $-23$ & $-21$ & $-19$ & $-21$ \\
$(c_L)^{TT}=(c_L)^{JJ}$&8&{\it I,II}&
$-25$ & $-24$ & $-19$ & $-21$ & $-22$ \\
$\half(c_L)^{(TJ)}$&24&{\it I,II,III,IV}&
$-25$ & $-24$ & $-19$ & $-21$ & $-22$ \\
$\half(c_L)^{(JK)}-\fr13\de^{JK}(c_L)^{TT}$&40&{\it I,II,III,IV}&
$-25$ & $-24$ & $-19$ & $-21$ & $-22$ \\
$\gt^{JT}-\fr i2\ep^{JKL}\gt^{KL}$&36&{\it I,II,III,IV,V,VI}&
$-25$ & $-24$ & $-19$ & $-21$ & $-22$ \\
$\half\gt^{(JK)}-\fr13\de^{JK}\gt^{LL}$&60&{\it I,II,III,IV,V,VI}&
$-25$ & $-24$ & $-19$ & $-21$ & $-22$ \\
$\Ht^J$&18&{\it I,II,III,IV,V}&
$-27$ & $-23$ & $-21$ & $-19$ & $-21$ \\
$(k_d)^{\la\ldots}$&var.\ &{\it I,II,III,IV,V,VI}&
$-27+2d$ & $-23-d$ & $-21+2d$ & $-19-2d$ & $-21-d$ \\
\hline
\hline
\end{tabular}
\caption{Experimental prospects.}
\end{table*}

We conclude with some comments 
about prospects for Lorentz- and CPT-violation searches
in the major types of experiments.
Table I provides a summary of the present situation.
The left-hand part of the table contains three columns
with information about coefficients for Lorentz violation. 
The first column lists 
combinations of coefficient matrices 
relevant to neutrino propagation, 
extracted from the general hamiltonian \rf{heff}
and separated according to rotation properties
into timelike ($T$) and spacelike ($J$) components 
in the Sun-centered frame.
The second column lists the maximum number 
of independent degrees of freedom (DOF) 
associated with each combination of coefficient matrices.
These numbers can be obtained by examining the form 
of Eq.\ \rf{heff} and using the symmetry properties 
in generation space listed in Eq.\ \rf{C}.
In certain specific models,
some of these degrees of freedom may be unobservable.
The third column displays the classes of signal 
that are relevant for each coefficient matrix,
using the nomenclature of the previous subsection.
The right-hand part of the table 
contains estimated attainable sensitivities,
classified according to each of five types of oscillation experiments.
Each entry in the table represents the
base-10 logarithm of the expected sensitivity 
to the corresponding coefficient for Lorentz violation.
The sensitivities shown in the table can be obtained 
by examination of Fig.\ \ref{expts}.
Given an experiment 
with maximum $L$ coverage of $L_{\rm max}$
and maximum $E$ coverage of $E_{\rm max}$,
the crude sensitivity $\si$ 
to a coefficient for Lorentz violation of dimension $1-d$
is taken to be
$\si \approx - \log L_{\rm max} - d \log E_{\rm max}$.
For simplicity in the presentation,
it is understood that the sensitivities 
listed for the dimension-one coefficients 
$a_L$, $H$ are measured in GeV.
The final row of the table contains a rough estimate
of sensitivities measured in GeV$^{(1-d)}$
to a generic coefficient $(k_d)_{\la\ldots}$ 
for a Lorentz-violating operator 
of nonrenormalizable dimension $n=d+3$.
Some caution is required in interpreting the latter numerical estimates
because the coefficients $(k_d)_{\la\ldots}$ 
are expected typically to be suppressed
by $d$-dependent powers of the Planck scale. 

The table confirms that Planck-scale sensitivities 
to Lorentz and CPT violation are attainable 
in all classes of experiment,
with the most sensitive cases 
potentially rivaling the best tests in other sectors of the SME.
Note that the estimated sensitivities
assume order-one measurements and therefore may underestimate 
the true attainable sensitivity in any specific experiment. 
Note also that a variety of experimental analyses
are needed to extract complete information on Lorentz and CPT violation,
with no single class of experiment
presently in a position to provide 
measurements of a complete set of coefficients.
In the remainder of this subsection,
we offer a few more specific remarks 
about each type of experiment.

\begin{trivlist}

\item
{\it Solar-neutrino experiments.}
The abundance and quality of the current solar-neutrino data 
make these experiments a promising avenue  
for Lorentz-violation searches.
The relatively large range of solar-neutrino energies 
suggests interesting information about spectral anomalies
might be obtained,
but complications introduced by matter effects 
are likely to make this practical 
only in relatively simple cases 
such as the FC model \rf{rim}.
Of the other classes of signals,
periodic variations and neutrino-antineutrino mixing 
may be the most relevant to solar neutrinos.
The periodic variations in observables 
would occur at multiples of $\Om_\oplus$,
appearing despite compensation for the flux variation
due to the eccentricity of the Earth's orbit.
Direct detection of any antineutrinos originating from the Sun 
would be evidence of $\nu\mix\nub$ mixing
and hence of possible Lorentz violation.

\item
{\it Atmospheric-neutrino experiments.}
Like solar neutrinos,
atmospheric neutrinos cover a relatively large region 
of $L$--$E$ space,
but complications from matter effects
hinder a general spectral-anomaly search.
However, 
Fig.\ \ref{expts2} shows 
that searches for atmospheric oscillations 
at the highest energies and largest distances 
could reveal oscillations absent in the usual solution,
thereby providing evidence for $L$--$E$ conflicts.
Atmospheric neutrinos originate from all directions,
so they are an ideal system for directional-dependence searches.
Not only are they sensitive to sidereal variations,
but also the directional capabilities of detectors like SK
make atmospheric neutrinos perhaps the most promising place 
to search for compass asymmetries.
Moreover,
since atmospheric data involve both neutrinos and antineutrinos 
of two species in comparable numbers,
it may be possible to address both $\nu\mix\nub$ mixing 
and the classic CPT tests \rf{cpt1} and \rf{cpt2}.

\item
{\it Reactor experiments.}
Nuclear reactors are good sources of $\nub_e$,
and they are therefore well suited 
to searches for $\nu\mix\nub$ mixing.
Since both the sources and the detectors in all these cases are fixed,
the experiments are also sensitive to sidereal variations,
and some may have additional sensitivity to compass asymmetries.
For example,
the reactor experiment KamLAND detected neutrinos 
from multiple reactors and different locations.
Experiments with multiple sources like this 
can analyze their data for compass asymmetries 
that depend on the direction to the various neutrino sources.

\item
{\it Short-baseline accelerator experiments.}
LSND already seems to suggest a positive 
$L$--$E$ conflict,
which will be tested by the forthcoming results 
of the MiniBooNE experiment.
Many of these short-baseline accelerator experiments
are especially interesting for signals based on $L$--$E$ conflicts
because they operate in a region of $L$--$E$ space
where the conventional mass scenario predicts no oscillations.
Sidereal variations can readily be sought by experiments  
such as CHORUS, KARMEN, MiniBooNE, NOMAD, and NuTeV,
since each has a fixed source and detector.
Note that the existing data from these experiments 
could in principle contain a positive signal for sidereal variations
because the published null results are based on an average over time.
The well-defined flavor content of the sources for these experiments 
may also offer sensitivity to $\nu\mix\nub$ signals 
and to the classic CPT test.
Some of these experiments,
such as MiniBooNE and NuTeV,
may be particularly sensitive to Lorentz violation
because they can switch from a predominately 
$\nu_\mu$ source to a predominately $\nub_\mu$ source.

\item
{\it Long-baseline accelerator experiments.}
Several future long-baseline accelerator-based experiments,
such as ICARUS, MINOS, and OPERA,
are planned to probe the GeV region of $L$--$E$ space
at distances of hundreds of kilometers,
and some results in this regime have already been reported by K2K.
These experiments can search for oscillations in $\nu_\mu$ 
obtained from meson decays,
and they are designed to test the atmospheric-oscillation hypothesis.
Nonetheless, 
$L$--$E$ conflicts are still possible:
a measurement of $\nu_\mu\to\nu_e$,
for example, 
would represent an $L$--$E$ conflict
because this oscillation is absent at these energies and distances 
in the conventional scenario with masses.
The data obtained can be also analysed for sidereal variations,
since in each case the source and detector are fixed.
Moreover, 
except for OPERA and ICARUS,
which are both 
part of the CERN Neutrinos to Gran Sasso (CNGS) project,
the beamline for each experiment points 
in a different direction.
This means each is expected to respond differently 
to rotation-invariance violations.
These experiments may also be able to address
$\nu\mix\nub$ mixing and the classic CPT signal,
since the flavor content of the beams is well known.

\item
{\it Other experiments.}
Experiments designed to search for neutrino properties 
other than oscillations can also address Lorentz violation.
To some extent,
most experiments are sensitive to sidereal variations
and compass asymmetries.
The other signals discussed in Sec.\ \ref{predictions}
are more unique to neutrino oscillations,
but analogous signatures are likely to arise in most cases.

One possible test of Lorentz invariance
involves a direct comparison of velocities
of supernova neutrinos and photons,
such as those from SN1987A
\cite{sn1987a,snnugamma},
which could be performed 
either by some of the experiments listed above 
or by neutrino telescopes. 
A similar method has been applied in the photon sector,
where the velocities of different polarizations are compared
\cite{km}.
Another method that could be adapted to the neutrino case 
is a simple pulse-dispersion analysis.
The energy dependence 
and the independent propagation of each $\heff$ eigenstate
imply that different components of the neutrino pulse
propagate at different velocities,
causing the pulse to spread.
For SN1987A, 
all the observed neutrinos arrived 
in a time interval of about $\de T \simeq 10$ s
and had energies $E\simeq 10-20$ MeV.
Since these neutrinos took roughly $T_0 \simeq 5\times 10^{12}$ s
to reach the Earth,
we can crudely estimate that the maximum difference in velocity 
across the $\De E\simeq 10$ MeV energy spread 
of the $\heff$ eigenstates
is $\de v/c\simeq\de T/T_0\simeq 2\times 10^{-12}$.
We can then make a simple dimensional estimate of the sensitivity 
of this method to various terms in $\heff$.
This suggests a sensitivity of about $\sqrt{200}$ eV to mass terms,
$2\times 10^{-14}$ GeV to dimension-one coefficients,
and $2\times 10^{-12}$ to dimensionless coefficients.
The mass estimate agrees 
with the result of a detailed analysis along these lines 
\cite{snnu}.

Lorentz violation may also be relevant to 
direct mass searches such as the proposed KATRIN experiment
\cite{katrin},
designed to measure directly the $\nu_e$ mass to better than 1 eV.
Within the currently accepted solution to the oscillation data,
a mass matrix with eV-scale masses but
mass-squared differences of $10^{-3}$ eV$^2$ and $10^{-5}$ eV$^2$ 
would be nearly degenerate.
This seems unlikely in light of the charged-lepton mass hierarchies.
However, 
suppose that the mass matrix is nearly diagonal
and that neutrino oscillations are primarily
or entirely due to Lorentz violation instead.
Then,
individual masses of eV order or greater  
may be present with little or no effect 
on the existing neutrino-oscillation data,
but they would produce a signal in the KATRIN experiment.

Another area of widespread interest
is the search for neutrinoless double-beta decay
\cite{bbdecay}.
This decay mode is an indicator of lepton-number violation,
which can result from Majorana-type couplings 
introduced by Majorana masses
or by gauge-violating coefficients for Lorentz violation.
Many of the null results of searches for neutrinoless double-beta decay
could therefore be reanalysed to yield constraints 
on certain types of Lorentz violation.

\end{trivlist}

\acknowledgments
This work was supported in part by the 
United States Department of Energy
under grant number DE-FG02-91ER40661
and by the 
National Aeronautics and Space Administration
under grant number NAG8-1770.


\appendix

\section{Effective hamiltonian}\label{hcalc}

This appendix presents some details for the derivation 
of the effective hamiltonian \rf{heff}.
We first perform a spinor decomposition of the hamiltonian 
in the mass-diagonal Majorana basis.
The result is then block diagonalized in the light-neutrino sector
and transformed into the original weak-interaction basis.
We remind the reader that 
generation indices in the mass-diagonal basis are
$\AA=1,\ldots, 6$ for $N=3$ neutrino species,
while the restriction to light neutrinos in this basis 
is represented by indices $\aa=1,2,3$.
Also,
in the flavor basis,
upper-case indices take the values $A=e,\mu,\ta,e^C,\mu^C,\ta^C$,
while lower-case ones span $a=e,\mu,\ta$.

\subsection{Spinor decomposition}

In this section,
we project the hamiltonian onto the massless spinor basis 
used in Eq.\ \rf{nup}.
This corresponds to choosing a convenient
$\vp$-dependent $\ga$-matrix basis
that allows us to write the equations of motion 
in terms of the $b$ and $d$ amplitudes.

Working in the mass-diagonal basis,
the hamiltonian is given by 
\bea
\cH_{\AA\BB}(\vp)&=&
\cH_{0\AA\BB}(\vp)+\de\cH_{\AA\BB}(\vp) ,
\nonumber \\
\cH_{0\AA\BB}(\vp)&=&
\ga^0(\vec\ga\cdot\vp+m_{(\AA )})\de_{\AA\BB}\ ,
\nonumber \\
\de\cH_{\AA\BB}(\vp)
&=&-\half(\ga^0\de\Ga^0\cH_0(\vp)+\cH_0(\vp)\ga^0\de\Ga^0)_{\AA\BB}
\nonumber \\
&&+\ga^0(\de\vec\Ga\cdot\vp+\de M)_{\AA\BB}\ .
\label{cH}\eea
It turns out to be useful to decompose 
$\Ga^\mu_{\AA\BB}$ and $M_{\AA\BB}$
in terms of $\ga$ matrices,
as in Eq.\ \rf{GaM}.
Therefore, we write
\bea
\Ga^\nu_{\AA\BB} &=&
\ga^0U_{\AA A}\ga^0\Ga^\nu_{AB}(U_{\BB B})^\dag
\nonumber \\
&=&\ga^\nu \de_{\AA\BB}
+ c^\mn_{\AA\BB}\ga_\mu
+ d^\mn_{\AA\BB}\ga_5\ga_\mu
\nonumber \\&&
+ e^\nu_{\AA\BB}
+ if^\nu_{\AA\BB}\ga_5
+ \half g^{\la\mn}_{\AA\BB}\si_{\la\mu}\ ,
\nonumber \\
M_{\AA\BB} &=&
\ga^0U_{\AA A}\ga^0M_{AB}(U_{\BB B})^\dag
\nonumber \\
&=&m_{\AA\BB}
+im_{5\AA\BB}\ga_5
\nonumber \\&&
+ a^\mu_{\AA\BB}\ga_\mu
+ b^\mu_{\AA\BB}\ga_5\ga_\mu
+ \half H^\mn_{\AA\BB}\si_\mn\ .\quad
\label{GaMp}\eea

We begin the spinor decomposition of the hamiltonian \rf{cH}
by considering the Lorentz-covariant terms.
The properties of the massless spinor basis imply 
that the only nonzero projections of the kinetic term are 
\bea
\lefteqn{
  u_{L,R}^\dag(\vp)(\ga^0\vec\ga\cdot\vp\ \de_{\AA\BB}) u_{L,R}(\vp)}
\nonumber \\
&=&-v_{R,L}^\dag(-\vp)(\ga^0\vec\ga\cdot\vp\ \de_{\AA\BB}) v_{R,L}(-\vp)
\nonumber \\
&=&|\vp|\de_{\AA\BB}\ ,
\eea
while the surviving projections of the mass term are
\bea
\lefteqn{
  u_{L,R}^\dag(\vp)(\ga^0 m_{(\AA)} \de_{\AA\BB}) v_{R,L}(-\vp)}
\nonumber \\
&=&\bar u_{L,R}(\vp)v_{R,L}(-\vp) m_{(\AA)} \de_{\AA\BB}\ 
\eea
and conjugates.
The quantities $\bar u_{L,R}(\vp)v_{R,L}(-\vp)$
are phases that can be chosen arbitrarily 
by changing the relative phase between
$u_{L,R}(\vp)$ and $v_{R,L}(-\vp)=C\bar u^T_{L,R}(-\vp)$.

For the spinor decomposition of the Lorentz-violating terms
in the hamiltonian \rf{cH},
we define the $2\times 2$ matrices
\bea
\lefteqn{
  \La_{\AA\BB}(\vp)=
  \La_{\BB\AA}^\dag(\vp)}
\nonumber \\
&=&\left( \begin{array}{c}
u_L^\dag(\vp) \\ u_R^\dag(\vp)
\end{array} \right)
\de\cH_{\AA \BB}(\vp)
\big( u_L(\vp) , u_R(\vp) \big) ,
\label{La} \\
\lefteqn{
  \tilde\La_{\AA \BB}(\vp)=
  -\tilde\La_{\BB \AA}^T(-\vp)}
\nonumber \\
&=&\left(\begin{array}{c}
u_L^\dag(\vp) \\ u_R^\dag(\vp)
\end{array}\right)
\de\cH_{\AA \BB}(\vp)
\big( v_R(-\vp) , v_L(-\vp) \big) .\quad
\label{tLa}
\eea
It can be shown that the mass-basis analogues of
the relations \rf{GaMC} are
$\Ga^\nu_{\AA\BB}=-C(\Ga^\nu_{\BB\AA})^TC^{-1}$
and $M_{\AA\BB}=C(M_{\BB\AA})^TC^{-1}$.
Note that this corresponds to $\cmat\to I$,
which reflects the Majorana nature of neutrinos in this basis.
These identities may then be used to
show that $C^\dag\ga^0\cH_{\AA\BB}(\vp)\ga^0C
=-[\cH_{\AA\BB}(-\vp)]^*$.
Finally, 
with the aid of the relation
$v_{R,L}(\vp)=C\bar u^T_{L,R}(\vp)$,
it follows that the remaining terms in the spinor decomposition 
are given in terms of $\La$ and $\tilde\La$ by
\bea
\lefteqn{-\tilde\La_{\AA\BB}^*(-\vp)}
\nonumber \\ && =
\left(\begin{array}{c}
v_R^\dag(-\vp) \\ v_L^\dag(-\vp)
\end{array}\right)
\de\cH_{\AA\BB}(\vp)
\big( u_L(\vp) , u_R(\vp) \big) , \\
\lefteqn{-\La_{\AA\BB}^*(-\vp)}
\nonumber \\ && =
\left(\begin{array}{c}
v_R^\dag(-\vp) \\ v_L^\dag(-\vp)
\end{array}\right)
\de\cH_{\AA\BB}(\vp)
\big( v_R(-\vp) , v_L(-\vp) \big) . \quad\quad
\eea
This implies that the $2\times 2$ matrices
$\La_{\AA\BB}$, $\tilde\La_{\AA\BB}$
determine the Lorentz-violating effects.

Combining the above results,
we obtain the spinor-decomposed hamiltonian 
appearing in Eq.\ \rf{we}:
\bea
\lefteqn{
H_{\AA\BB}(\vp)=H_{\BB\AA}^\dag(\vp) }
\nonumber \\
&=&\de_{\AA\BB}\left(\begin{array}{cc}
|\vp|
&
m_{(\AA)}\et(\vp)
\\
-m_{(\AA)}\et^*(-\vp)
&
-|\vp|
\end{array}
\right)
\nonumber \\
&&\quad
+\left(\begin{array}{cc}
\La_{\AA\BB}(\vp)
&
\tilde\La_{\AA\BB}(\vp)
\\
-\tilde\La_{\AA\BB}^*(-\vp)
&
-\La_{\AA\BB}^*(-\vp)
\end{array}
\right) ,
\label{H}
\eea
where $\et$ is the $2\times 2$ diagonal matrix of phases
$\et(\vp)=-\et(-\vp)
={\rm diag}[\bar u_L(\vp) v_R(-\vp),
\bar u_R(\vp) v_L(-\vp)]$.

We seek an explicit expression for $\La_{\AA\BB}$.
The next subsection shows that 
the effects of $\tilde\La_{\AA\BB}$ are subleading order,
so we concentrate here on the projections in $\La_{\AA\BB}$,
which involve the spinors $u_L$ and $u_R$.
It is useful first to find expressions for the quantities
$\bar u_\al \{1,\ga_5,\ga^\mu,\ga_5\ga^\mu,\si^\mn\}u_\be$,
where $\al,\be=L,R$.
We obtain the following nonzero results:
\bea
\bar u_\al \ga^\mu u_\be&=&p^\mu\de_{\al\be}/|\vp| ,
\nonumber \\
\bar u_\al \ga_5\ga^\mu u_\be&=&S_\al p^\mu\de_{\al\be}/|\vp| ,
\nonumber \\
\bar u_L \si^\mn u_R&=&
(\bar u_R \si^\mn u_L)^*
\nonumber \\
&=& i\sqrt{2}(p^\mu(\ep_+)^\nu-p^\nu(\ep_+)^\mu)/|\vp| ,\quad
\label{spinpr}\eea
where 
$S_L=1$, 
$S_R=-1$, 
$p^\mu=(|\vp|;\vp)$,
and $(\ep_+)^\mu$ satisfies the relations \rf{ep}.
With these results and Eqs.\ \rf{cH} and \rf{GaMp},
we can extract the projections of $\de\cH$
onto $u_L$ and $u_R$:
\begin{widetext}
\beq
\La_{\AA\BB}=
\fr{1}{|\vp|}
\left(\begin{array}{cc}
[(a+b)^\mu p_\mu-(c+d)^\mn p_\mu p_\nu]_{\AA\BB} &
-i\sqrt{2}p_\mu(\ep_+)_\nu[
g^{\mn\si}p_\si-H^\mn]_{\AA\BB} \\
i\sqrt{2}p_\mu(\ep_+)^*_\nu[
g^{\mn\si}p_\si-H^\mn]_{\AA\BB} &
[(a-b)^\mu p_\mu-(c-d)^\mn p_\mu p_\nu]_{\AA\BB}
\end{array}\right) .
\eeq
\end{widetext}
In this expression,
we neglect off-diagonal terms entering as mass 
multiplied by coefficients for Lorentz violation,
since in most situations these
terms are suppressed relative to those above.

\subsection{Block diagonalization}

The above spinor decomposition of the hamiltonian 
is independent of the specific neutrino mass spectrum.
To make further progress,
we adopt the scenario described at the beginning
of Section \ref{phenom}
and restrict attention to ultrarelativistic dynamics
in the subspace of light neutrinos,
spanned by the $\aa$ indices.
The hamiltonian is then dominated by
the diagonal kinetic term in Eq.\ \rf{H}.
The upper and lower diagonal blocks of this term have opposite sign, 
so they differ by an amount large compared to 
both mass and Lorentz-violating terms.
This in turn implies that standard perturbation techniques 
to remove the off-diagonal blocks can be applied.
As a result,
terms in the off-diagonal blocks of Eq.\ \rf{H} 
appear at second order in the block-diagonalized form.
One consequence is that the leading-order mass contribution 
appears at second order,
whereas certain forms of Lorentz violation appear already at first order.
This feature can ultimately be traced to the $\ga$-matrix structure 
of the Lorentz-covariant portion of the theory.

Provided the conditions 
$m_{(\aa )},|\La_{\aa \bb}| |\tilde\La_{\aa \bb}| \ll |\vp|$
are satisfied,
the block diagonalization of Eq.\ \rf{H}
can proceed through the perturbative construction 
of an appropriate unitary matrix ${\cal U}$.
First,
write ${\cal U}$ in the form
${\cal U}=I+\ep^{(1)}+\ep^{(2)}+\ldots$,
where $\ep^{(n)}$ is of $n$th order in
the dimensionless small quantities
$m_{(\aa )}/|\vp|$,
$\La_{\aa \bb}/|\vp|$, and
$\tilde\La_{\aa \bb}/|\vp|$.
The block-diagonal hamiltonian resulting from this transformation
can be expanded in a similar fashion:
\bea
H_{\aaa \bbb}&=&{\cal U}_{\aaa \aa}H_{\aa \bb}{\cal U}_{\bbb \bb}^\dag
\nonumber \\
&=&H^{(0)}_{\aaa \bbb}+H^{(1)}_{\aaa \bbb}+H^{(2)}_{\aaa \bbb}+\cdots ,
\eea
where each $H^{(n)}_{\aaa \bbb}$
is $n$th order in small quantities.
The zeroth-order term $H^{(0)}_{\aaa \bbb}$ is the usual kinetic term,
which is already block diagonal.
The first-order term $H^{(1)}_{\aaa \bbb}$
can be made block diagonal by an appropriate choice of $\ep^{(1)}$.
A suitable leading-order transformation is
\beq
\ep^{(1)}_{\aaa \bb}=
\fr{\de_{\aaa \aa}}{2|\vp|}
\left(\begin{array}{cc}
0 & \hat\ep_{\aa\bb}(\vp) \\
\hat\ep^*_{\aa\bb}(-\vp) & 0 \\
\end{array}\right) ,
\eeq
where
\beq
\hat\ep_{\aa\bb}(\vp)=
m_{(\aa )}\de_{\aa \bb}\et(\vp)
+\tilde\La_{\aa\bb}(\vp) .
\eeq
Using $\ep^{(1)}$ and $H^{(2)}_{\aaa \bbb}$,
which depends on both $\ep^{(1)}$ and  $\ep^{(2)}$,
we can find $\ep^{(2)}$ 
and then continue iteratively to arbitrary order.

Under the transformation $\cal U$,
the hamiltonian restricted to light neutrinos
may be written
\beq
H_{\aaa \bbb}=
\left(\begin{array}{cc}
h_{\aaa \bbb}(\vp) & 0 \\
0 & -h_{\aaa \bbb}^*(-\vp)
\end{array}\right) .
\eeq
Calculating $\cal U$ to second order in small quantities
yields the second-order hamiltonian
\beq
h_{\aaa \bbb}(\vp)=
\de_{\aaa \aa}\de_{\bbb \bb}
\left[
\left(|\vp|
+\frac{1}{2|\vp|}m_{(\aa)}^2
\right)
\de_{\aa \bb}
+\La_{\aa \bb}(\vp)
\right] .
\label{h}
\eeq
This expression neglects terms 
that are second order in coefficients for Lorentz violation
and terms that enter as the product of $m_{(\aa)}/|\vp|$
with $\tilde\La$.
The latter terms constitute subleading-order corrections
under the reasonable assumption that $\La$ and $\tilde\La$ are
comparable in size.

While formally the two bases related by $\cal U$ are different,
in practice this difference is of little consequence.
Our main goal is to determine oscillation probabilities.
The effects of $\cal U$ appear in the mixing matrix and
therefore modify the amplitudes of oscillations.
However, 
since $\cal U$ is close to the identity,
the basis change produces only tiny and unobservable changes 
in oscillation amplitudes.
It therefore suffices in practice to assume ${\cal U} = I$
for purposes of the basis transformation,
corresponding to ignoring the difference between the
$\aa$ and $\aaa$ indices.
Similar arguments apply to the field redefinition
relating $\nu$ and $\ch$.
This also underlies the validity of assuming 
unitarity mixing matrices in the conventional case with neutrino mass,
even though the submatrix $V_{\aa a}$ is only approximately unitary.
In contrast,
the diagonalization of $h$ in Eq.\ \rf{h} 
can introduce arbitrary amounts of mixing.

The above description in the mass-diagonal basis 
completely determines the neutrino dynamics,
but in practical situations a description 
in the weak-interaction basis is more useful.
This requires the transformation of $h_{\aa\bb}$ 
to the original flavor basis.

The first step in implementing the desired transformation
is to determine the relation between the coefficients
in Eq.\ \rf{GaM} and those in Eq.\ \rf{GaMp}.
In terms of the unitary matrix $V_{\AA A}$,
we find
\bea
c^\mn_{\AA\BB}&=&\Re  V_{\AA A}V^*_{\BB B}(c+d)^\mn_{AB} ,
\nonumber \\
d^\mn_{\AA\BB}&=&i\Im  V_{\AA A}V^*_{\BB B}(c+d)^\mn_{AB} ,
\nonumber \\
e^\nu_{\AA\BB}&=&i\Im V_{\AA A}V_{\BB B}[(e+if)^\nu\cmat]_{AB} ,
\nonumber \\
if^\nu_{\AA\BB}&=&\Re V_{\AA A}V_{\BB B}[(e+if)^\nu\cmat]_{AB} ,
\nonumber \\
\half g^{\la\mn}_{\AA\BB}&=&
 \Re V_{\AA A}V_{\BB B}\half(g^{\la\mn}\cmat)_{AB}
\nonumber \\
 &&-\Im V_{\AA A}V_{\BB B}\frac14\ep^{\la\mu\rh\si}
 ({g_{\rh\si}}^\nu\cmat)_{AB} ,
\nonumber \\
m_{\AA\BB}&=&\Re V_{\AA A}V_{\BB B}[(m+im_5)\cmat]_{AB}
 \equiv m_{(\AA)}\de_{\AA\BB} ,
\nonumber \\
im_{5\AA\BB}&=&i\Im V_{\AA A}V_{\BB B}[(m+im_5)\cmat]_{AB}
 \equiv0 ,
\nonumber \\
a^\nu_{\AA\BB}&=&i\Im V_{\AA A}V^*_{\BB B}(a+b)^\nu_{AB} ,
\nonumber \\
b^\nu_{\AA\BB}&=&\Re V_{\AA A}V^*_{\BB B}(a+b)^\nu_{AB} ,
\nonumber \\
\half H^\mn_{\AA\BB}&=&
  i\Im V_{\AA A}V_{\BB B}\half(H^\mn\cmat)_{AB}
\nonumber \\
  &&+i\Re V_{\AA A}V_{\BB B}\frac14\ep^{\mn\rh\si}
  (H_{\rh\si}\cmat)_{AB} .
\label{coeffs}
\eea
Note that all the coefficients in the mass-diagonal basis
are either pure real or pure imaginary,
reflecting the Majorana nature of neutrinos in this basis.
Using this equation,
we obtain
\bea
\lefteqn{
  [(a+b)^\mu p_\mu-(c+d)^\mn p_\mu p_\nu]_{\aa\bb}}
\nonumber \\
&&=
[(a+b)^\mu p_\mu-(c+d)^\mn p_\mu p_\nu]_{ab}
V_{\aa a} V^*_{\bb b}\ ,
\nonumber\\
\lefteqn{
  [(a-b)^\mu p_\mu-(c-d)^\mn p_\mu p_\nu]_{\aa\bb}}
\nonumber \\
&&=[-(a+b)^\mu p_\mu-(c+d)^\mn p_\mu p_\nu]^*_{ab}
V^*_{\aa a} V_{\bb b}\ ,
\nonumber\\
\lefteqn{
  -i\sqrt{2}p_\mu(\ep_+)_\nu[
  g^{\mn\si}p_\si-H^\mn]_{\aa\bb}}
\nonumber \\
&&=-i\sqrt{2} p_\mu (\ep_+)_\nu[
(g^{\mn\si}p_\si-H^\mn)\cmat]_{ab}
V_{\aa a} V_{\bb b}\ ,
\nonumber\\
\lefteqn{
  i\sqrt{2}p_\mu(\ep_+)^*_\nu[
  g^{\mn\si}p_\si-H^\mn]_{\aa\bb}}
\nonumber \\
&&=i\sqrt{2}p_\mu(\ep_+)^*_\nu[
(g^{\mn\si}p_\si+H^\mn)\cmat]^*_{ab}
V^*_{\aa a} V^*_{\bb b}\ ,\quad\quad
\eea
using the assumption that the submatrix
$V_{\aa a}$ is unitary.

Within a standard seesaw mechanism,
the right-handed Majorana-mass matrix $R$ 
appearing in Eq.\ \rf{usumass} is large, 
$|R| \gg |L|,|D|$.
Calculating the matrix $V_{AB}$ 
at leading order in small mass ratios
$|L|/|R|$ and $|D|/|R|$ produces the identity
\beq
m_{(\aa)}\de_{\aa\bb}=V_{\aa a}V_{\bb b}(\ml)_{ab} ,
\eeq
where $\ml= L - DR^{-1}D^T$,
and hence the relation
\beq
m_{(\aa)}^2\de_{\aa\bb}
=V_{\aa a}V^*_{\bb b}(\ml \ml^\dag)_{ab}
=V^*_{\aa a}V_{\bb b}(\ml \ml^\dag)^*_{ab} .
\eeq
Combining results yields the desired form,
\bea
\lefteqn{
\left[
\left(|\vp|+\frac{1}{2|\vp|}m_{(\aa)}^2\right)
\de_{\aa \bb}
+\La_{\aa \bb}(\vp)
\right]}
\nonumber \\
&&=\left(\begin{array}{cc}
V_{\aa a} & 0 \\ 0 & V^*_{\aa a}
\end{array}\right)
(\heff)_{ab}
\left(\begin{array}{cc}
V^*_{\bb b} & 0 \\ 0 & V_{\bb b}
\end{array}\right) ,
\eea
where $\heff$ is given in
Eq.\ \rf{heff}.


\phantom{xxxxxxxxxxxxxxxxxxxxxxxxx}

\section{Minimal SME terms}\label{smeterms}

Restricting attention to 
the coefficients $(c_L)^\mn_{ab}$, $(a_L)^\mu_{ab}$,
which are contained in the minimal gauge-invariant SME,
effectively decouples neutrinos and antineutrinos
and produces vanishing transition probabilities 
\rf{pnb} and \rf{pbn}.
This appendix describes a useful parametrization 
of these coefficients.

Each coefficient matrix for Lorentz violation
can be parametrized with three eigenvalues 
and a constant unitary matrix.
We define 
\beq
(c_L)^\mn=
(\hat U^\mn)^\dag
\left(\begin{array}{ccc}
(c_L)^\mn_{(1)} & 0 & 0 \\
0 & (c_L)^\mn_{(2)} & 0 \\
0 & 0 & (c_L)^\mn_{(3)}
\end{array}\right)
\hat U^\mn 
\eeq
for each coefficient matrix $(c_L)^\mn$,
and
\beq
(a_L)^\mu=
(\hat U^\mu)^\dag
\left(\begin{array}{ccc}
(a_L)^\mu_{(1)} & 0 & 0 \\
0 & (a_L)^\mu_{(2)} & 0 \\
0 & 0 & (a_L)^\mu_{(3)}
\end{array}\right)
\hat U^\mu 
\eeq
for each coefficient matrix $(a_L)^\mu$.
The unitary diagonalizing matrices
$\hat U^\mn$, $\hat U^\mu$
are chosen so that 
if there is only a single nonvanishing coefficient matrix
then $\Ueff$ in Eq.\ \rf{diaheff}
takes the block-diagonal form
\beq
\Ueff=
\left(\begin{array}{cc}
\hat U & 0 \\
0 & \hat U^* \\
\end{array}\right) .
\eeq
The reader is warned that the above decomposition
is frame dependent,
so neither the eigenvalues nor the mixing matrices
behave as tensors under observer Lorentz transformations.
We therefore advocate restricting this type of decomposition
to the standard Sun-centered celestial equatorial frame.

Adopting a CKM-like decomposition of the $\hat U$ matrices,
we denote mixing angles and phases
associated with each $(c_L)^\mn$ by
$\th_{12}^\mn$, 
$\th_{13}^\mn$, 
$\th_{23}^\mn$,
and 
$\de^\mn$, 
$\be_1^\mn$, 
$\be_2^\mn$.
Similarly, 
for each $(a_L)^\mu$ we write 
$\th_{12}^\mu$, 
$\th_{13}^\mu$, 
$\th_{23}^\mu$,
and 
$\de^\mn$, 
$\be_1^\mu$, 
$\be_2^\mu$.
The $\hat U$ matrices may then be written explicitly 
in the form
\begin{widetext}
\bea
\hat U^\mn&=&
\left[\begin{array}{ccc}
c^\mn_{12}c^\mn_{13} &
-s^\mn_{12}c^\mn_{23}-c^\mn_{12}s^\mn_{23}s^\mn_{13}e^{-i\de^\mn} &
s^\mn_{12}s^\mn_{23}-c^\mn_{12}c^\mn_{23}s^\mn_{13}e^{-i\de^\mn} \\
s^\mn_{12}c^\mn_{13} &
c^\mn_{12}c^\mn_{23}-s^\mn_{12}s^\mn_{23}s^\mn_{13}e^{-i\de^\mn} &
-c^\mn_{12}s^\mn_{23}-s^\mn_{12}c^\mn_{23}s^\mn_{13}e^{-i\de^\mn} \\
s^\mn_{13}e^{i\de^\mn} &
s^\mn_{23}c^\mn_{13} &
c^\mn_{23}c^\mn_{13} \\
\end{array}\right]
\left[\begin{array}{ccc}
1 & 0 & 0 \\
0 & e^{i\be_1^\mn} & 0 \\
0 & 0 & e^{i\be_2^\mn} \\
\end{array}\right] , \\
\hat U^\mu&=&
\left[\begin{array}{ccc}
c^\mu_{12}c^\mu_{13} &
-s^\mu_{12}c^\mu_{23}-c^\mu_{12}s^\mu_{23}s^\mu_{13}e^{-i\de^\mu} &
s^\mu_{12}s^\mu_{23}-c^\mu_{12}c^\mu_{23}s^\mu_{13}e^{-i\de^\mu} \\
s^\mu_{12}c^\mu_{13} &
c^\mu_{12}c^\mu_{23}-s^\mu_{12}s^\mu_{23}s^\mu_{13}e^{-i\de^\mu} &
-c^\mu_{12}s^\mu_{23}-s^\mu_{12}c^\mu_{23}s^\mu_{13}e^{-i\de^\mu} \\
s^\mu_{13}e^{i\de^\mu} &
s^\mu_{23}c^\mu_{13} &
c^\mu_{23}c^\mu_{13} \\
\end{array}\right]
\left[\begin{array}{ccc}
1 & 0 & 0 \\
0 & e^{i\be_1^\mu} & 0 \\
0 & 0 & e^{i\be_2^\mu} \\
\end{array}\right] ,
\eea
\end{widetext}
where
$s^\mn_{ab}=\sin\th^\mn_{ab}$,
$c^\mn_{ab}=\cos\th^\mn_{ab}$,
$s^\mu_{ab}=\sin\th^\mu_{ab}$,
and
$c^\mu_{ab}=\cos\th^\mu_{ab}$.

In the conventional massive-neutrino analysis,
the $\be$ matrix of phases can be absorbed 
into the amplitudes $b_a(t;\vp)$ and $d_a(t;\vp)$, 
so these phases are normally unobservable and can be neglected.
However, 
in the present context,
only one set of $\be$ phases may be absorbed in this fashion.
The presence of multiple coefficient matrices for Lorentz violation
implies that they cannot typically be neglected.

Neutrino oscillations are insensitive to terms 
in the effective hamiltonian that are proportional 
to the identity.
Consequently, 
only two eigenvalue differences 
for each coefficient matrix for Lorentz violation
contribute to oscillation effects.
Also, 
each coefficient matrix is associated with
three mixing angles and three phases.
It follows that
the maximum number of gauge-invariant degrees of freedom
that enter into neutrino oscillations in the minimal SME alone
is 16$\times$8 for $c_L$ and 4$\times$8 for $a_L$,
for a total of 160.
However,
some of these are unobservable.
The 8 trace components $\et_\mn (c_L)^\mn$ are Lorentz invariant,
and both these and
the 6$\times$8-component antisymmetric piece of $(c_L)^\mn$
are absent in the leading-order hamiltonian \rf{heff}.
This leaves 104 leading-order degrees of freedom 
in $a_L$ and $c_L$,
in agreement with the numbers listed in Table I.
For the minimal SME, 
one set of $\be$ phases is also unobservable,
which reduces the total number of degrees of freedom to $102$.



\begin{thebibliography}{99}
\bibitem{fn1}
The precise definitions of the Sun-centered frame and
of a standard Earth-based frame 
appropriate for terrestrial experiments
are given in Ref.\ \cite{km},
along with the form of the transformation relating them.
\bibitem{fn2}
This example exhibits maximal mixing at $\Th=90^\circ$,
but it is straightforward to generate 
$\Th$-dependent examples 
with maximal mixing at $\Th=0^\circ, 180^\circ$.
See Ref.\ \cite{nu}.
\bibitem{fn3}
Note that this behavior may also result
from direction dependence or
from resonances in $\heff$ at intermediate energy scales.
\end{thebibliography}
\end{document}